\documentclass[%
 reprint,
showpacs,
nofootinbib,
 amsmath,amssymb,
 aps,
prd,
]{revtex4-1}

\usepackage{graphicx}
\usepackage{dcolumn}
\usepackage{bm}
\usepackage{textcomp}
\usepackage{amsmath}
\usepackage{amssymb}
\usepackage{comment}
\usepackage{subfigure}
\newcommand{\ra}[1]{\renewcommand{\arraystretch}{#1}}
\newcommand{\GeV}{\text{\,GeV}}
\newcommand{\TeV}{\text{\,TeV}}

\newcommand{\keV}{\text{\,keV}}
\newcommand{\eV}{\text{\,eV}}

\newcommand{\MPl}{M_\text{Pl}}
\newcommand{\cm}{\, \text{cm}}
\newcommand{\Hz}{\, \text{Hz}}

\newcommand{\pd}[2]{\frac{\partial #1}{\partial #2}}
\newcommand{\half}{\frac{1}{2}}

\begin{document}

\preprint{SU-ITP-14/12}

\title{Searching for dilaton dark matter with atomic clocks}

\author{Asimina Arvanitaki}
 \email{aarvanitaki@perimeterinstitute.ca}
 \affiliation{Perimeter Institute for Theoretical Physics,
 Waterloo, Ontario, N2L 2Y5, Canada}
\author{Junwu Huang}%
 \email{curlyh@stanford.edu}
\author{Ken Van Tilburg}
 \email{kenvt@stanford.edu}
\affiliation{%
 Stanford Institute for Theoretical Physics, Department of Physics,
 Stanford University, Stanford, CA 94305, USA
}%

\date{\today}

\begin{abstract}
We propose an experiment to search for ultralight scalar dark matter (DM) with dilatonic interactions.  Such couplings can arise for the dilaton as well as for moduli and axion-like particles in the presence of CP violation. Ultralight dilaton DM acts as a background field that can cause tiny but coherent oscillations in Standard Model parameters such as the fine structure constant and the proton-electron mass ratio. These minute variations can be detected through precise frequency comparisons of atomic clocks. Our experiment extends current searches for drifts in fundamental constants to the well-motivated high-frequency regime.  Our proposed setups can probe scalars lighter than $10^{-15}\eV$ with discovery potential of dilatonic couplings as weak as $10^{-11}$ times the strength of gravity, improving current equivalence principle bounds by up to 8 orders of magnitude. We point out potential $10^4$ sensitivity enhancements with future optical and nuclear clocks, as well as possible signatures in gravitational wave detectors. Finally, we discuss cosmological constraints and astrophysical hints of ultralight scalar DM, and show they are complimentary to and compatible with the parameter range accessible to our proposed laboratory experiments.
\end{abstract}

\maketitle




\section{Introduction}
\label{sec:intro}

With the Higgs discovery at the Large Hadron Collider (LHC), the Standard Model (SM) is now complete. However successful it may be, the particle content of the SM only accounts for about $5\%$ of the energy content of our universe. Motivated by the hierarchy problem and the WIMP miracle, we had associated the scale of another $27\%$ of our world with the TeV scale, and expected the LHC and direct detection experiments to shed light on the nature of dark matter (DM). Their null results to date diminish the connection of new physics with the electroweak scale and deepen the mystery of the scale and properties of DM.

The largest component of the cosmos offers the biggest mystery of all. The cosmological constant (CC) challenges our notion of Naturalness by more than 60 orders of magnitude. There is no experimental evidence for physics beyond the SM at the relevant energy scale; theories that try to predict the CC value stumble on this fact, even if we ignore the theoretical inconsistencies that usually plague those models. This leaves us with one known framework where a complete picture of the cosmos can be embedded: the string landscape. The many vacua of string theory could accommodate the smallness of the CC due to environmental selection. 

Even though such a framework cannot be tested directly at low energies, it gives rise to a variety of indirect signatures. The topologically complex manifolds that are required to produce the large number of vacua also imply the existence of many particles: axions, moduli, photons, dilatons, or even entire hidden sectors. These particles have widely varying properties and viable mass ranges, and it is possible that several may significantly contribute to the DM of our universe. From the above multitude of candidates, we would like to entertain the possibility that the DM is composed of an ultralight boson with dilatonic couplings. Contrary to fermionic DM candidates, the misalignment mechanism for light bosons produces a non-thermal, cold component of DM and opens the parameter space to the sub-keV mass range. The couplings of these bosons to SM particles could be greatly suppressed, circumventing the usual collider and direct detection constraints. The scalar mass can be protected by shift symmetry or the smallness of the couplings, or perhaps it could be tuned to be small, just like the Higgs mass and the CC.

For masses well below $1\eV$, bosonic DM in our galaxy has densities that exceed $\lambda_{\text{dB}}^{-3}$, where $\lambda_{\text{dB}}$ is the de Broglie wavelength of the particle. In this case, the scalar DM exhibits coherence and behaves like a wave with amplitude $\sim {\sqrt{\rho_\text{DM}}}/{m_\text{DM}}$ and coherence time $2 \pi (m_{\text{DM}} v_{\text{vir}}^2)^{-1}$, where $v_{\text{vir}}$ is the virial velocity of DM in our galaxy \cite{calcaxion}. This coherence offers a new avenue for detecting DM and can be used to enhance the reach of laboratory experiments.

In this paper, we focus on the non-derivative coupling of such DM candidates to the SM:
\begin{equation}
\label{eqn:coupling-gen}
\frac{\phi}{M^*}\mathcal{O}_{\text{SM}},
\end{equation}
where $\mathcal{O}_{\text{SM}}$ denotes terms in the SM Langrangian. The scale $M^*$ can be many orders of magnitude above the Planck scale.
Such couplings exist for the dilaton and moduli, as well as for axions in the presence of CP violation.  When such a particle is the DM, its scalar coupling to the SM will cause coherent oscillations of fundamental constants, such as the fine structure constant or the proton-electron mass ratio, at a frequency set by the DM mass. These will in turn cause oscillations in the energy levels of atoms and can thus have a clear signature in devices that set our current standard for time: \textit{atomic clocks}. 

Rapid advances in frequency metrology---in particular the technology of mode-locked lasers and the self-referencing frequency comb \cite{holzwarth2000optical,PhysRevLett.84.5102,udem2002optical, Cundiff:2003zz,diddams2010evolving}---have exhibited high precision that can be sensitive to these minute oscillations.  The long-term stability of the best $^{133}$Cs atomic clocks at the $10^{-16}$ level has recently been eclipsed by that of optical clocks, which are approaching stabilities of $10^{-18}$.  Since the energies of atomic transitions in clocks based on different elements have varying dependencies on the proton-electron mass ratio and the fine-structure constant, their comparison can reveal the effects of a scalar dark matter field with couplings as in Eq.~\eqref{eqn:coupling-gen}.  Changes in the frequency of a microwave or optical transition can now be measured to incredible precision with a femtosecond frequency comb produced by mode-locked lasers.  The technology and methods involved in our experimental proposal have already been demonstrated in labs around the world; we propose a new way of using these tools to search for new fundamental physics improving current searches by many orders of magnitude.

We start by defining the framework for couplings of the schematic form presented in Eq.~\eqref{eqn:coupling-gen} in Sec.~\ref{sec:model}. In Sec.~\ref{sec:setup}, we outline a simplified version of our experimental proposal. We quantify the sensitivity of our setup to the scalar coupling of DM in Sec.~\ref{sec:sensitivity}, and compare the reach to existing limits from fifth-force searches and EP tests in Sec.~\ref{sec:EPtests}. Potential signatures in gravitational-wave observatories are presented in Sec.~\ref{sec:gravwaves}. We discuss the cosmology and the astrophysical constraints on ultralight scalars in Sec.~\ref{sec:cosmo}. Finally, we conclude in Sec.~\ref{sec:conclusions}. A discussion on possible ultraviolet completions of dilaton DM is presented in Appendix~\ref{sec:dpmech}.

\section{Dilaton Couplings of Dark Matter}
\label{sec:model}

We consider an ultralight singlet scalar field that makes up all---or an $\mathcal{O}(1)$ fraction---of the local dark matter density, and focus on its possible scalar couplings with the Standard Model through higher-dimensional operators.  Adopting the conventions of \cite{Damour:2010rp}, the relevant low-energy couplings can be written as:
\begin{align}
S &= \int d^4x \sqrt{|g|}~\left\lbrace \frac{1}{2} \partial_\mu \phi \partial^\mu \phi - V(\phi) + \mathcal{L}_\text{SM} + \mathcal{L}_{\phi} \right\rbrace \nonumber\\
\mathcal{L}_{\phi} &= \kappa \phi \bigg[+\frac{d_e}{4e^2}F_{\mu\nu}F^{\mu\nu} -\frac{d_g \beta_3}{2g_3}G^A_{\mu\nu}G^{A\mu\nu} \nonumber\\
&\hspace{4mm} - d_{m_e} m_e \bar{e} e - \sum_{i = u,d} (d_{m_i} + \gamma_{m_i} d_g)m_i \bar{\psi}_i \psi_i \bigg], \label{eq:couplings}
\end{align}
where $\kappa \equiv \frac{\sqrt{4\pi}}{\MPl}$, $\beta_3$ is the QCD beta function, and $\gamma_{m_i}$ are the anomalous dimensions of the $u$ and $d$ quarks. We parametrize the scalar potential as 
\begin{align}\label{eq:scalarpot}
V(\phi) = \frac{1}{2} m_\phi^2 \phi^2 + \frac{1}{3} a_\phi \phi^3 + \frac{1}{4}\lambda_\phi \phi^4.
\end{align}
The effective couplings in Eq.~\eqref{eq:couplings} are ubiquitous in models involving very light scalars in semi-hidden sectors.  Most notably, the QCD axion and other axion-like particles also have scalar couplings in the presence of CP violation, in addition to their usual pseudoscalar couplings. For the QCD axion there is a non-zero scalar coupling to the quarks \cite{Pospelov:1997uv,Pospelov:2001ys,Pospelov:2005pr}:
\begin{align}
 \left(\frac{10^{-16}}{f_a}\right)\lesssim \kappa d_{\hat m}\lesssim  \left(\frac{3\cdot 10^{-11}}{f_a}\right),\label{eq:axion}
\end{align}
where $f_a$ is the axion decay constant, and $d_{\hat{m}}$ is the coupling to the symmetric combination of the quark masses:
\begin{align}
d_{\hat{m}} \equiv \frac{d_{m_d} m_d +d_{m_u} m_u}{m_d + m_u} \label{eq:dmhat}.
\end{align}
The upper bound on the scalar coupling is set by neutron electric dipole moment searches, while the lower bound is set by the amount of CP violation in the Standard Model. Note that there is a large theoretical uncertainty in the exact value of this coupling. We also expect $d_g$, $d_{m_e}$ and $d_e$ couplings to be radiatively generated for the QCD axion. 

Another way in which non-derivative couplings of a light scalar can appear is through a Higgs portal. In the model of \cite{Piazza:2010ye}, a super-renormalizable coupling to the Higgs $\mathcal{L} \supset A \phi H^\dagger H$ induces couplings of an ultralight scalar to the SM, suppressed by a factor ${A v_\text{ew}}/{ m_h^2}$ relative to the scalar couplings of the Higgs. However, the simplest model requires $A < \sqrt{2 \lambda_h} m_\phi$ to avoid an unstable direction in the scalar potential.  Hence the expected couplings to e.g.~the fermion masses are $d_{m_i} \lesssim 10^{-13} \left(\frac{m_\phi}{10^{-18}\eV} \right)$, probably too small to detect with current technology, as we will show in Sec.~\ref{sec:sensitivity}. For the remainder of the paper, we shall focus on the phenomenology of Eqs.~\eqref{eq:couplings}~\&~\eqref{eq:scalarpot}, and postpone a discussion of ultraviolet embeddings of this dilaton-like theory until appendix \ref{sec:dpmech}.

With the action in Eq.~\eqref{eq:couplings}, a tiny mass $m_\phi$, and sufficiently weak couplings---negligible self-interactions and $d_i \ll 1$---the history of the scalar field on scales larger than $m_\phi^{-1}$ is well-approximated by a background field that starts oscillating when $H \sim m_\phi$ and whose energy density redshifts as that of ordinary cold, pressureless DM.   Its present-day behaviour can then be described as the solution to its equation of motion:
\begin{align}\label{eq:phisol}
\phi(t, \vec{x}) = \phi_0 \cos(m_\phi t - \vec{k}_\phi\cdot \vec{x} + \dots).
\end{align}
We take its current energy density $\rho_{\phi} = \half m_\phi^2 \phi_0^2$ to equal the local DM density $\rho_{\text{DM}} \approx 0.3 \GeV/\cm^3$, and assume a wave vector given by the virial velocity: $|\vec{k}_\phi| \simeq m_\phi v_\text{vir}$ with $v_\text{vir} \approx 10^{-3}$.  With these assumptions, we obtain the fractional amplitude of $\phi$ relative to the reduced Planck scale: 
\begin{align}\label{eq:phiamp}
\kappa \phi_0 = \frac{\sqrt{8\pi \rho_\phi}}{m_\phi \MPl} = 6.4\cdot 10^{-13} \left(\frac{10^{-18}\eV}{m_\phi}\right)F^{1/2},
\end{align}
where $F \equiv \rho_\phi / \rho_\text{DM}$ is the fractional contribution of $\phi$ to the local DM density.

The classic probes of the scalar couplings in Eq.~\eqref{eq:couplings} are fifth-force experiments, and searches for deviations from the weak equivalence principle (EP).   Current technology can set limits on the $d_i$ down to the $10^{-5}$ level, as we will discuss in Sec.~\ref{sec:EPtests}.  
With one extra assumption---that the energy density in the $\phi$ field makes up the DM---other phenomenological signatures come into play, such as temporal and spatial variation of mass ratios and gauge couplings. Writing $\kappa \phi$ as the field normalized to the reduced Planck mass $\MPl/\sqrt{4\pi}$, the above definitions are such that
\begin{align} \label{eq:oscobs}
\pd{\ln  \Lambda_3}{(\kappa \phi)} &= d_g, ~~ 
\pd{\ln m_i(\Lambda_3)}{(\kappa \phi)} = d_{m_i}, ~~ \pd{ \ln \alpha }{(\kappa \phi)} = d_e,
\end{align}
where $\Lambda_3$ is the QCD confinement scale and $\alpha$ is the fine-structure constant.  Hence if the local value of the $\phi$ field changes, the effective masses and gauge couplings of fundamental particles change as well. The mass of the proton $m_p$ changes primarily due to the change of the QCD confinement scale $\Lambda_3$ via the $d_g$ coupling, and subdominantly due to the change of the quark masses $m_{u,d}$ via the symmetric $d_{\hat{m}}$ coupling defined in Eq.~\eqref{eq:dmhat}. 
Comparing this with Eq.~\eqref{eq:oscobs}, we quickly see that experiments with extreme sensitivity to changes in $m_{p} / m_e$ or $\alpha$ can conceivably constrain the parameters $d_i$ to sub-unity levels.

\section{Concept and experimental setup}
\label{sec:setup}

The goal of our proposal is to search for dark-matter-induced oscillations in the frequency ratios of certain atomic transitions, which can be electronic, hyperfine, or nuclear in nature.  Since the effects will be small, we need stable sources of light with a narrow line width, as well as a way to measure the frequency of pairs of such lines precisely.  Cavity-stabilized lasers locked to atomic transitions---atomic clocks---provide such monochromatic light with fractional stabilities down to $10^{-18}$ \cite{2013arXiv1305.5869H,Bloom13}; mode-locked lasers generating femtosecond pulses can be used as measurement ``tools" to compare their frequencies \cite{udem2002optical,Cundiff:2003zz}.

The basic observation is that the energies $f_\text{A}$ of different atomic transitions (labeled by A) can have varying scalings with certain mass ratios and the fine structure constant:
\begin{align}\label{eq:coeffpowers}
f_\text{A} \propto \left(\frac{\mu_\text{A}}{\mu_b}\right)^{\zeta_\text{A}} (\alpha)^{\xi_\text{A}+2}
\end{align}
where $\mu_\text{A}$ is the nuclear magnetic moment of nucleus A, and $\mu_b$ is the Bohr magneton.  The ratio $\mu_\text{A}/\mu_b$ is linearly proportional to $m_e/m_p$ and the orbital and spin $g$-factors of the nucleus.
The exponent $\zeta_\text{A}$ is 1 for hyperfine transitions, and 0 for optical transitions. We have factored out a conventional exponent of 2 in the $\alpha$ dependence due to the scaling of the Rydberg constant $R_\infty \propto \alpha^2$.  

Via Eqs.~\ref{eq:phisol}~\&~\ref{eq:oscobs}, the dark matter causes minute, coherent oscillations in $\mu_\text{A}/\mu_b$ and $\alpha$:
\begin{align}\label{eq:constantosc}
\left(\frac{\mu_\text{A}}{\mu_b} \right) &\simeq \left(\frac{\mu_\text{A}}{\mu_b}\right)_0 \left[1 + \left(d_{m_e} - d_g + M_\text{A} d_{\hat{m}} \right) \kappa \phi(t) \right]; \nonumber \\
\alpha &= \alpha_{0} \left[ 1 + d_e \kappa \phi(t) \right].
\end{align}
The scaling of $\mu_\text{A}/\mu_b$ with the light quark masses, as parametrized by the $M_\text{A} d_{\hat{m}}$ term, has been estimated for several relevant nuclei in \cite{FlambaumCs}; for e.g.~$\text{A} = {^{133}\text{Cs}}$, we have $M_A \approx + 0.07$. From Eqs~\ref{eq:coeffpowers}~\&~\ref{eq:constantosc}, we thus obtain fractional variations in the frequency ratio of atomic transitions A and B:
\begin{align}\label{eq:freqratosc}
\frac{\delta \left({f_\text{A}}/{f_\text{B}}\right)}{ {f_\text{A}}/{f_\text{B}}}\simeq \Big[\zeta_\text{A} \left(d_{m_e} - d_g + M_\text{A} d_{\hat{m}} \right)  + \Delta\xi_\text{AB} d_e \Big] \kappa \phi(t),
\end{align}
with $\Delta \xi_\text{AB} \equiv \xi_A - \xi_B$. For example, if A is a hyperfine microwave transition and B is an electronic optical transition, $\zeta_\text{A} = 1$, yielding sensitivity to $d_g$, $d_{m_e}$, and $d_{\hat{m}}$.\footnote{Actually, such a comparison only constrains a (non-compact) region in the two-dimensional parameter space $\lbrace\left(d_{m_e} - d_g + M_\text{A} d_{\hat{m}}\right) ,d_e\rbrace$. With clock comparisons of different pairs of atomic transitions, one can disentangle this degeneracy.}
If A and B are two different optical transitions, $\zeta_\text{A} = 0$, but $\Delta \xi_\text{AB}$ will be an $\mathcal{O}(1)$ number (see Table~\ref{tab:elementlist}), yielding sensitivity to $d_e$.  We shall take B to be an optical transition throughout the paper.

Before we describe the setup of our experiment, we first review the generation of an optical frequency comb as well as the workings of an optical clock laser in Sec.~\ref{sec:comb}.  A description of how the ratio $f_\text{A}/f_\text{B}$ is measured follows in Sec.~\ref{sec:measurement}, in the case where A and B are both electronic transitions in the optical frequency range of $10^{14}$--$10^{15} \Hz$ (the modification to a microwave-optical comparison is straightforward). We postpone a discussion of nuclear transitions to Sec.~\ref{sec:future}. We will omit many technical details; the reader is referred to \cite{rosenband2008frequency,Lea:2008zz,Cundiff:2003zz} for more complete descriptions of these techniques.  

\begin{figure}[t]\begin{center}
\subfigure[]{\includegraphics[width=0.48\textwidth]{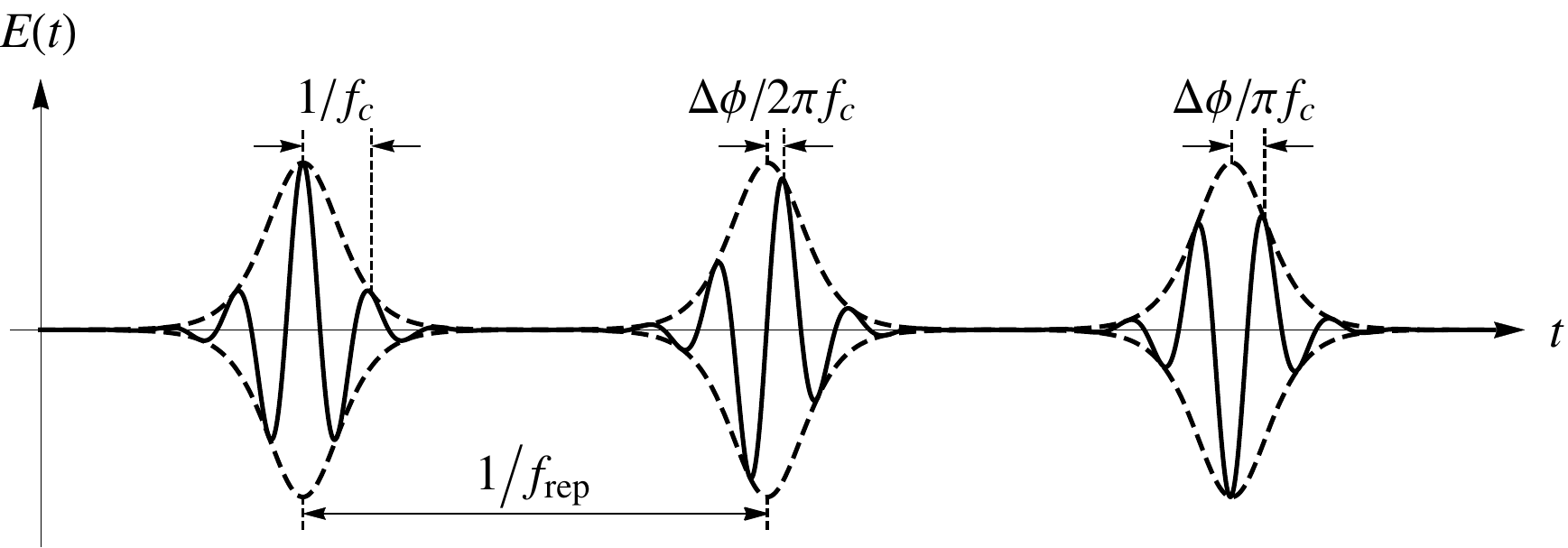}\label{fig:comba}}
\subfigure[]{\includegraphics[width=0.48\textwidth]{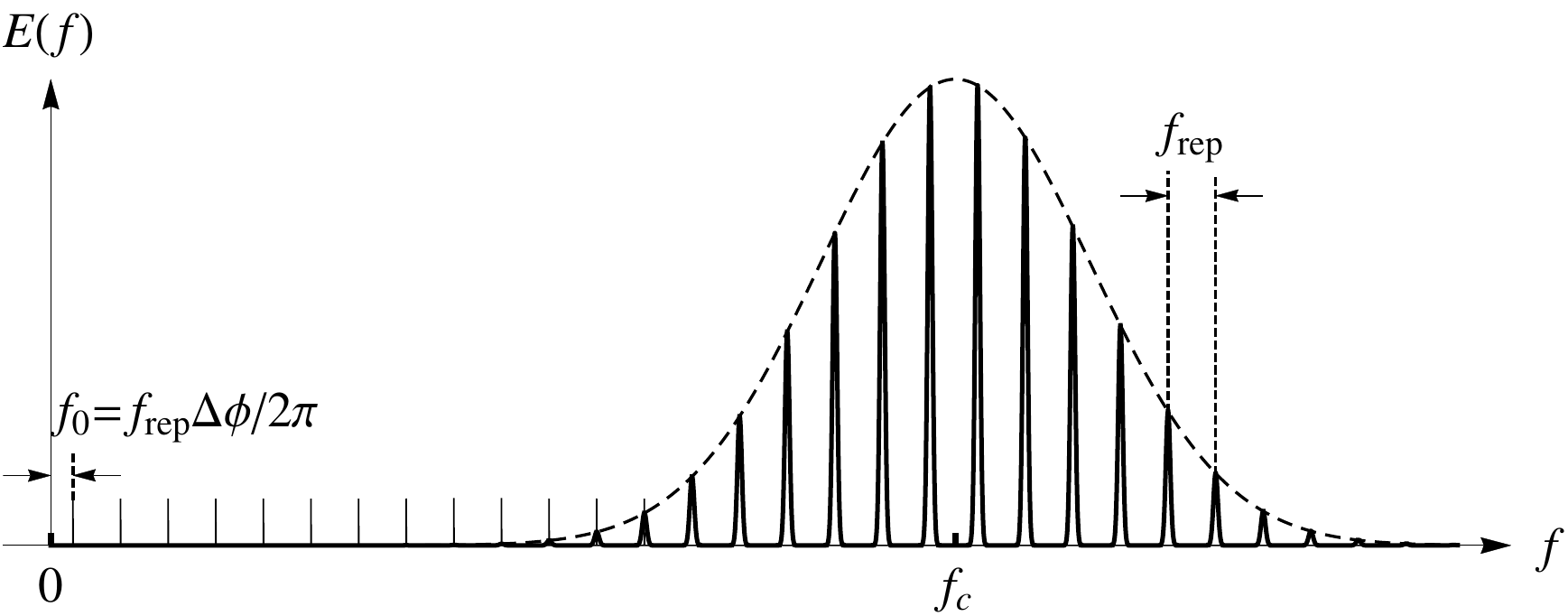}\label{fig:combb}}
\caption{(Color online) Diagrams of the (a) time and (b) frequency domains of a femtosecond laser with a time-evolving carrier-envelope phase $\phi_\text{ce}$, repetition rate $f_\text{rep}$, and carrier frequency $f_\text{c}$.  The comb in (b) is a Fourier transform of the pulse train in (a). In practice, $f_\text{c} / f_\text{rep} \sim 10^6$ but here we have taken it to be $\mathcal{O}(10)$ for illustrative purposes.}\label{fig:comb}
\end{center}
\end{figure}

\subsection{Frequency comb and atomic clock laser}\label{sec:comb}
It was first realized in \cite{baklanov1977narrow,PhysRevLett.38.760} that a regularly spaced train of short pulses (in the time domain) corresponds to a ``comb" in the frequency domain---a superposition of regularly spaced narrow lines.\footnote{See \cite{udem2002optical, Cundiff:2003zz,diddams2010evolving} for a more rigorous treatment and elaborate review including references; here we give a heuristic derivation.}
Development of mode-locked Ti:sapphire lasers has made the production of few-femtosecond-wide pulse trains possible.  To understand how a frequency comb is formed, first consider a single pulse.  Its frequency spectrum will be the Fourier transform of its envelope function, centered at the optical frequency of its carrier.  The width of the spectrum is inversely proportional to the temporal width of the pulse envelope.  For a train of pulses at fixed repetition rate $f_\text{rep}$, the Fourier expansion will be dominated by modes of \textit{discrete} frequencies separated by $f_\text{rep}$ (only for these can there be constructive interference), with largest amplitudes still near $f_\text{c}$ (see Fig.~\ref{fig:comb}).  A complication arises from the fact that the phase velocity of the optical carrier differs from the group velocity of the pulses in the gain medium of the femtosecond laser, causing a change $\Delta \phi_\text{ce}$ of the carrier-envelope phase between two pulses of magnitude $\Delta \phi_\text{ce} = 2\pi(v^{-1}_\text{group} - v^{-1}_\text{phase}) L_\text{cavity} f_\text{c} \mod 2\pi$, where $L_\text{cavity} = c / f_\text{rep}$ is the length of the cavity used for the femtosecond laser (see Fig.~\ref{fig:comb}).  In the frequency domain, this carrier-envelope phase shift manifests itself as a rigid frequency shift $f_0 = f_\text{rep} \Delta \phi_\text{ce} / 2\pi$ of the spectrum \cite{Cundiff:2003zz}.
The frequency comb lines are thus located at
\begin{align}\label{eq:comblines}
f_n = f_0 + n f_\text{rep}.
\end{align}
The pulse repetition rate $f_\text{rep}$ (and thus also $f_0$) of a femtosecond laser is typically in the microwave frequency band ($\sim 10^{8} \Hz$), meaning that the lines of an optical comb with significant power occur at very large $n$.  Measurement of $f_\text{rep}$ can be done with a fast photodiode \cite{jones2000carrier}, while measuring and stabilizing $f_0$ has also become a standard technique, in particular when the pulses are short enough so the comb spans at least an octave in frequency so ``self-referencing" becomes feasible \cite{Cundiff:2003zz}.  The stability of combs based on Ti:sapphire lasers has been demonstrated to be better than $8\cdot 10^{-20}$ \cite{ma2007frequency}, so Eq.~\eqref{eq:comblines} constitutes a near-perfect ``ruler'' in frequency space.

\begin{figure}[t]\begin{center}
\includegraphics[width=0.48\textwidth]{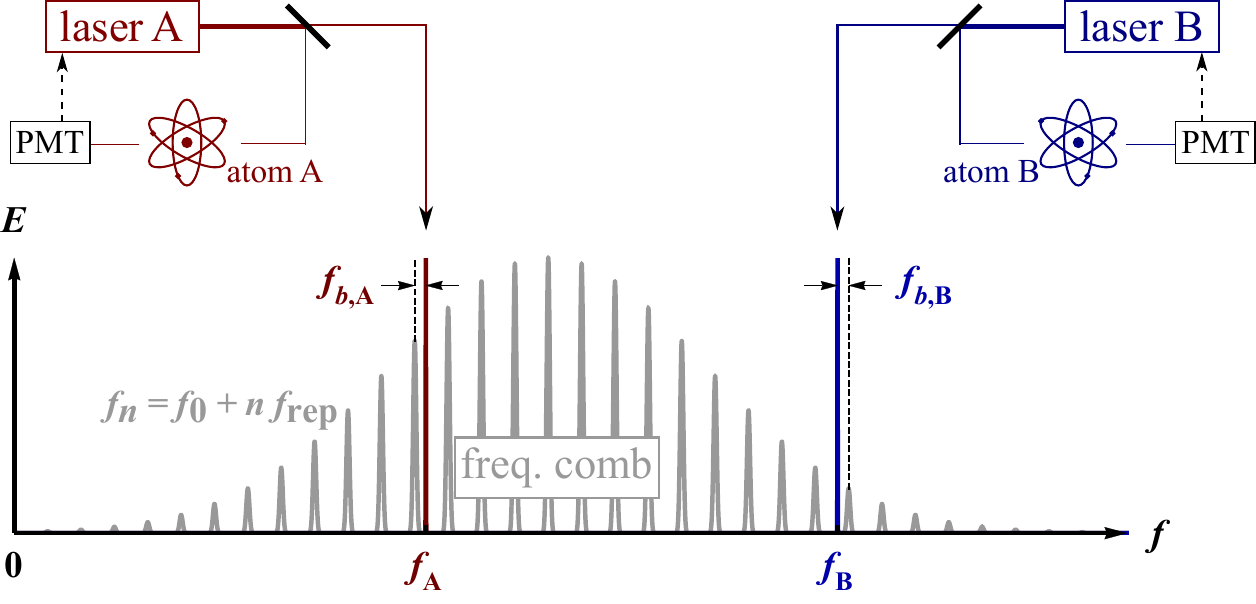}
\caption{(Color online) Experimental setup of a frequency comparison of two optical lines.  One clock laser based on a ``red'' transition $f_\text{A}$ and another based on a ``blue" transition $f_\text{B}$.  Light from both clock lasers is superposed with that from a frequency comb. Measurement of the beat frequencies $f_{b,\text{A}}$ and $f_{b,\text{B}}$, and comb frequencies $f_0$ and $f_\text{rep}$, provides the necessary information for a frequency ratio measurement. }\label{fig:setup}
\end{center}
\end{figure}

An optical clock laser is a cavity-stabilized laser whose frequency is locked to the energy of an electronic transition inside an atom.  In practice, the continuous-wave laser source is incident on a trapped, laser-cooled ion or on a sample of free-falling cold atoms, while a photomultiplier tube (PMT) measures resonance absorption of the laser light at the desired electronic transition energy (see Fig.~\ref{fig:setup}).   The laser source has frequency-determining components that keep it stabilized to the cavity and locked on resonance with the transition in the ion/atoms through a feedback mechanism. Care is taken to eliminate fluctuations in the length of the stabilizing cavity, as well as noise sources in the transition energy of the single ion or the sample of cold atoms. In Table~\ref{tab:elementlist}, we list demonstrated short- and long-term stabilities of clock lasers based on a wide range of elements, alongside corresponding references to the state-of-the-art setups which contain ample discussion on sources of noise and instability in atomic clock lasers. In summary, a setup such as the one depicted in Fig.~\ref{fig:setup} can output ultranarrow laser light with frequencies $f_\text{A,B}$ of stability $\sim 10^{-15} \Hz^{-1/2}$, where we can take $f_\text{A,B}$ to precisely match energy level differences in ion/atom samples A and B.

\begin{center}
\begin{table*}[ht]
\ra{1.4}
  \begin{tabular*}{0.9\textwidth}{@{\extracolsep{\fill}} l | l c c c c c @{}}
  \hline \hline
    Species & Transition & $\lambda~(\mathrm{nm})$ & Short $\left(\frac{10^{-15}}{\sqrt{\text{Hz}}}\right)$ & Long $\left(10^{-18}\right)$  & $\zeta_\text{A}$& $\xi_\text{A} $  \\ \hline 
    ${\mathrm{{}^{133} Cs}}$~\cite{parker2010long} & hyperfine & $ 3.3 \cdot 10^7 $ & $2\cdot 10^2 $ & $360$ & 1 & $2.83$\\ 
    ${\mathrm{{}^{199} Hg^+}}$~\cite{rosenband2008frequency} & $\mathrm{5d^{10} 6s \, {}^2 S_{\frac{1}{2}}\leftrightarrow5d^9 6s^2 \, {}^2 D_{\frac{5}{2}}}$ & $ 282$ & $2.8$ & $19$ & 0 & $-3.19$ \\ 
    ${\mathrm{{}^{171} Yb^+}}$~\cite{huntemann2012high} & $\mathrm{4f^{14} 6s \, {}^2 S_{\frac{1}{2}} \leftrightarrow 4f^{13} 6s^2 \,{}^2 F_{\frac{7}{2}}}$ & $ 467$ & $2.0$& $71 $ & 0 &$-5.30$ \\ 
    ${\mathrm{{}^{27} Al^+}}$~\cite{chou2010frequency} & $\mathrm{3s^2 \,{}^1 S_0 \leftrightarrow 3s3p \,{}^3 P_0}$ & $ 267$ & $2.8 $& $8.6$ & 0 & $0.008$ \\
    ${\mathrm{{}^{88} Sr^+}}$~\cite{PhysRevLett.109.203002} & $\mathrm{5s \,{}^2 S_{\frac{1}{2}} \leftrightarrow 4d \, {}^2 D_{\frac{5}{2}}}$ & $ 674$ & $16$ & $25$ & 0 & $0.43$ \\
    ${\mathrm{{}^{171} Yb}}$~\cite{2013arXiv1305.5869H} & $\mathrm{6 s^2 \,{}^1 S_0 \leftrightarrow 6s6p \,{}^3 P_0}$ & $ 578$ & $0.32$ & $1.6$ & 0 & $0.31$ \\
    ${\mathrm{{}^{87} Sr}}$~\cite{Bloom13} & $\mathrm{5 s^2 \,{}^1 S_0 \leftrightarrow 5s5p \,{}^3 P_0}$ & $ 698$ & $0.34$ & $6.4$ & 0 &  $0.06$ \\ 
     ${\mathrm{{}^{162} Dy}}$~\cite{Leefer:2013waa} & $\mathrm{4 f^{10} 5d 6 s \leftrightarrow 4 f^{9} 5d^2 6 s}$ & $ 4.0 \cdot 10^8 $ & $4.0 \cdot 10^6$ & - & 0 & $8.5 \cdot 10^6$ \\ 
      ${\mathrm{{}^{164} Dy}}$~\cite{Leefer:2013waa} & $\mathrm{4 f^{9} 5d^2 6 s \leftrightarrow 4 f^{10} 5d 6 s }$ &  $ 1.3 \cdot 10^9$ & $1.3 \cdot 10^7$ & - & 0 & $-2.6 \cdot 10^6$ \\ 
    ${\mathrm{{}^{229m} Th^{3+}}}$~\cite{PhysRevLett.108.120802} & nuclear & $ \sim 1.6 \cdot 10^2 $ & $\sim 1 $ & $\sim 1 $ & - & $\sim 10^4$ \\ \hline \hline
  \end{tabular*}
  \caption{Transition type \& wavelength, and short- \& long-term stabilities of current state-of-the-art microwave and optical atom clocks and the planned thorium nuclear clock.  Variation of the line frequency with changes in the ratio $\mu_\text{A}/\mu_b$ and fine structure constant $\alpha$ are parametrized with $\zeta_\text{A}$ and $\xi_\text{A}$ as in Eq.~\eqref{eq:coeffpowers}.}\label{tab:elementlist}
\end{table*}
\end{center}

\subsection{Measurement of clock frequency ratios}\label{sec:measurement}
With two optical clock lasers and a frequency comb, the corresponding two atomic transition energies can be compared \cite{rosenband2008frequency,Lea:2008zz}.  Light from both clock lasers can be transported through fibers and superposed on that of a frequency comb, incident on a fast photodiode.  If $f_\text{A}$ and $f_\text{B}$ are included in the range of the comb, they will both be ``close" to lines $n_\text{A}$ and $n_\text{B}$ in the comb (see Fig.~\ref{fig:setup}), giving rise to beating patterns at microwave beat frequencies $f_{b,\text{A}}$ and $f_{b,\text{B}}$, respectively.  The measured frequency ratio is:
\begin{align}\label{eq:freqratiomeas}
\left.\frac{f_\text{A}}{f_\text{B}}\right|_\text{expt} = \frac{f_0 + n_\text{A} f_\text{rep} + f_{b,\text{A}}}{f_0 + n_\text{B} f_\text{rep} + f_{b,\text{B}}}.
\end{align}
Note that the measured frequencies on the right-hand side ($f_0$, $f_\text{rep}$, $f_{b,\text{A}}$, $f_{b,\text{B}}$) of Eq.~\eqref{eq:freqratiomeas} are all microwave frequencies (by construction).  They can all be referenced to the \textit{same} microwave frequency standard (hydrogen maser or Cs-fountain clock).  It follows that the stability of the optical clocks is the only limiting factor on the precision of the left-hand side of Eq.~\eqref{eq:freqratiomeas}, because the stability of the microwave frequency reference standard cancels out in the frequency ratio. Other noise sources such as general-relativistic time delay fluctuations coming from e.g.~changes in the gravitational potential due to Earth's motion in the gravitational field of the Sun (with peak changes of $\mathcal{O}(10^{-10})$ and inverse year frequency) are common for both clocks, and thus also cancel out in the ratio.

This setup can be modified to perform a comparison of a hyperfine line $f_\text{A}$ in the microwave regime to a line $f_\text{B}$ in the optical regime \cite{Stalnaker:2007}. A hydrogen maser can \textit{directly} measure the frequency $f_\text{A}$ of the clock based on the hyperfine transition, and still simultaneously determine $f_\text{B}$ with the help of a frequency comb as above:
\begin{align}\label{eq:freqratiomeas2}
\left.\frac{f_\text{A}}{f_\text{B}}\right|_\text{expt} = \frac{f_\text{A}}{f_0 + n_\text{B} f_\text{rep} + f_{b,\text{B}}}
\end{align}
where again all quantities on the right-hand side are microwave frequencies referenced to the same hydrogen maser.  Typically, the precision of the left-hand side of Eq.~\eqref{eq:freqratiomeas2} is limited by the short-term stability of the hyperfine atomic clock.

In Table~\ref{tab:elementlist}, we provide a summary of commonly used atomic clocks based on the hyperfine cesium transition and various electronic transitions, as well as a proposed nuclear clock based on thorium (see Sec.~\ref{sec:nuclear}).  The frequencies of various ``clock transitions" scale differently with changes in the magnetic moments and fine-structure constant, as shown by the $\zeta_\text{A}$ and $\xi_\text{A}$ coefficients of Eq.~\eqref{eq:coeffpowers} in the last two columns of Table~\ref{tab:elementlist}. This variation in $\xi_\text{A}$ mainly comes from a complex overall effect of spin-orbit couplings and many-body effects in the electron cloud, which have been calculated and listed in \cite{PhysRevA.59.230,Lea:2008zz}. In two isotopes of dysprosium, these effects conspire to yield near-degenerate electronic levels separated by an energy splitting in the microwave regime, offering competitive sensitivity to changes in $\alpha$ despite a worse fractional frequency stability \cite{Leefer:2013waa}.  We also list the demonstrated short- and long-term stability of clocks based on these transitions.

\section{Sensitivity}
\label{sec:sensitivity}

In this section, we quantify the reach of our proposed clock comparison experiments.  The basic modus operandi is to interrogate two atomic clocks A and B and measure their frequency ratio $f_\text{A}/f_\text{B}$ after averaging for a time $\tau_1$, and repeat this measurement at regular time intervals $\Delta \tau$ for a total integration time $\tau_\text{int}$. (We will assume minimal down-time, such that $\tau_1 \simeq \Delta \tau$.) The output of this procedure is a discrete time series of $f_\text{A}/f_\text{B}$ with a total $\tau_{\text{int}} / \Delta \tau$ number of points.  Dark-matter-induced oscillations in the frequency ratio would show up as an isolated peak in the discrete Fourier transform (DFT) of this time series, at a (monochromatic) frequency $f_\phi = m_\phi/2\pi$. If there is such a monochromatic peak in the DFT, one could determine the magnitude of some linear combination of the $\lbrace d_i \rbrace$ times the square root of the fractional abundance $\sqrt{\rho_\phi/\rho_\text{DM}}$ via Eqs.~\eqref{eq:phisol},~\eqref{eq:phiamp}~\&~\eqref{eq:constantosc} and the strength of the signal; a non-observation of a peak in the DFT would set constraints (depending on the noise).  In what follows, we estimate the detection reach for the couplings $\lbrace d_g, d_{\hat{m}}, d_{m_e}, d_e \rbrace$ at unity signal-to-noise ratio ($\text{SNR} = 1$), assuming only one coupling dominates.

\subsection{Microwave-optical clock comparison}\label{sec:hyperfineoptical}

The type of setup described in Sec.~\ref{sec:setup} for a microwave-optical transition frequency comparison has been performed before, most recently in \cite{Fortier:2007} by comparing a $^{133}$Cs-fountain atomic clock with a $^{199}$Hg$^+$ optical clock, and also in \cite{Peik:2004} with $^{171}$Yb instead of $^{199}$Hg$^+$.  These experiments were primarily sensitive to linear drifts in fundamental constants: they placed limits on e.g.~$\frac{\dot{\alpha}}{\alpha}$.  Unless the scalar field responsible for these variations is much lighter than the run time of the experiments (a few years, corresponding to a DM mass of $\sim$ $10^{-22}\eV$), this is not an optimized analysis for looking for scalar field dark matter: the expected phenomenology is \textit{coherent oscillations} in mass ratios and the fine structure constant.

We suggest monitoring the frequency ratio $f_{\text{Cs}}/f_{\text{Sr}^+}$ at a $1 \Hz$ sampling rate and $1\text{\,s}$ averaging time per measurement ($\Delta \tau \simeq \tau_1 \approx 1\,\text{s}$). It does not matter which optical clock is used for the comparison, as long as it has a better short-term stability than a Cs clock.  We picked $^{88}\text{Sr}^+$ because it has a relatively long wavelength for an optical clock (so the optical synthesis is easier), and because $\xi_\text{Cs} - \xi_{\text{Sr}^+} \approx 2.4$ is relatively small (see Table~\ref{tab:elementlist}), minimizing the obfuscating effect of the $d_e$ coupling in Eq.~\eqref{eq:freqratosc}. At such short interrogation times, the uncertainty $\sigma_1$ of the frequency ratio measurement is dominated by the (in)stability of the Cs fountain \cite{Fortier:2007}, at $2 \cdot 10^{-13} \Hz^{-1/2}$ \cite{Stalnaker:2007}.\footnote{In other words, the fractional stability of a Cs-fountain atomic clock is $2 \cdot 10^{-13} /\sqrt{\tau_1 / \text{s}}$ where $\tau_1$ is the interrogation/averaging time for a single measurement.} 
Expected coherence times of the dark matter oscillations are of order $\tau_\text{coh} \simeq 2 \pi (m_\phi v^2)^{-1}$ with $v \approx 10^{-3}$ \cite{calcaxion}, so we expect to boost the sensitivity $\sigma_1$ of a single measurement by a factor of $\beta = \left(\min\lbrace \tau_\text{int}, \tau_\text{coh} \rbrace / \Delta \tau \right)^{1/2}$---the square root of the number of coherent measurements.

\begin{figure}[t]
\begin{center}
\includegraphics[width=0.48\textwidth]{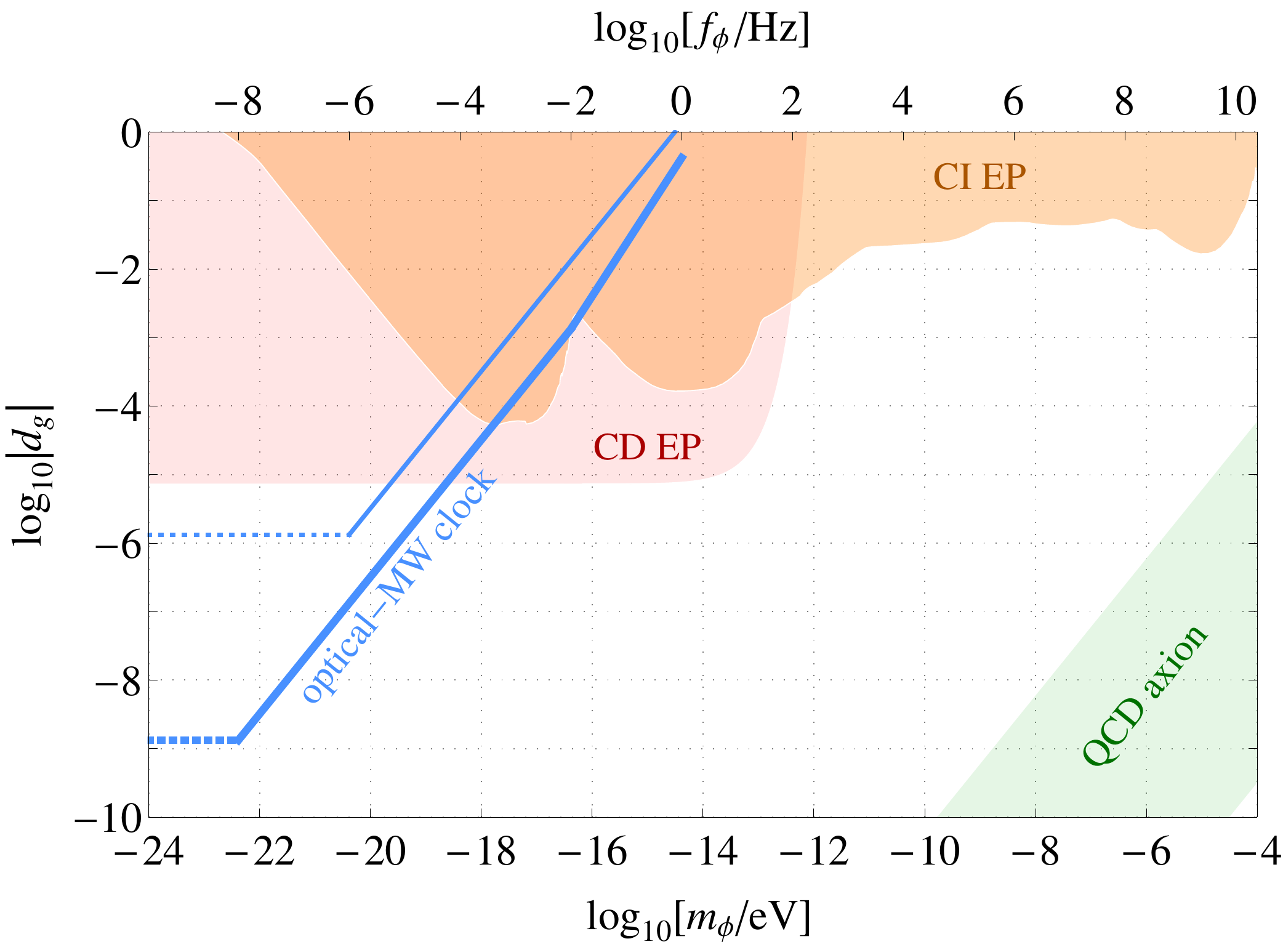}
\caption{(Color online) Sensitivity to $d_g$ as a function of the scalar dark matter mass $m_\phi = 2\pi f_\phi$ with the microwave-optical clock comparison experiment described in the text, for $\tau_\text{int} = 10^6,10^8\,s$ (thin blue, thick blue), assuming $\phi$ makes up all of the local dark matter.  Regions excluded by composition-dependent (CD EP) and -independent (CI EP) equivalence principle tests are colored in red and orange, respectively, assuming $d_g \gg d_{m_i},d_e$.  }\label{fig:dgbound}
\end{center}
\end{figure}

\begin{figure}[t]
\begin{center}
\includegraphics[width=0.48\textwidth]{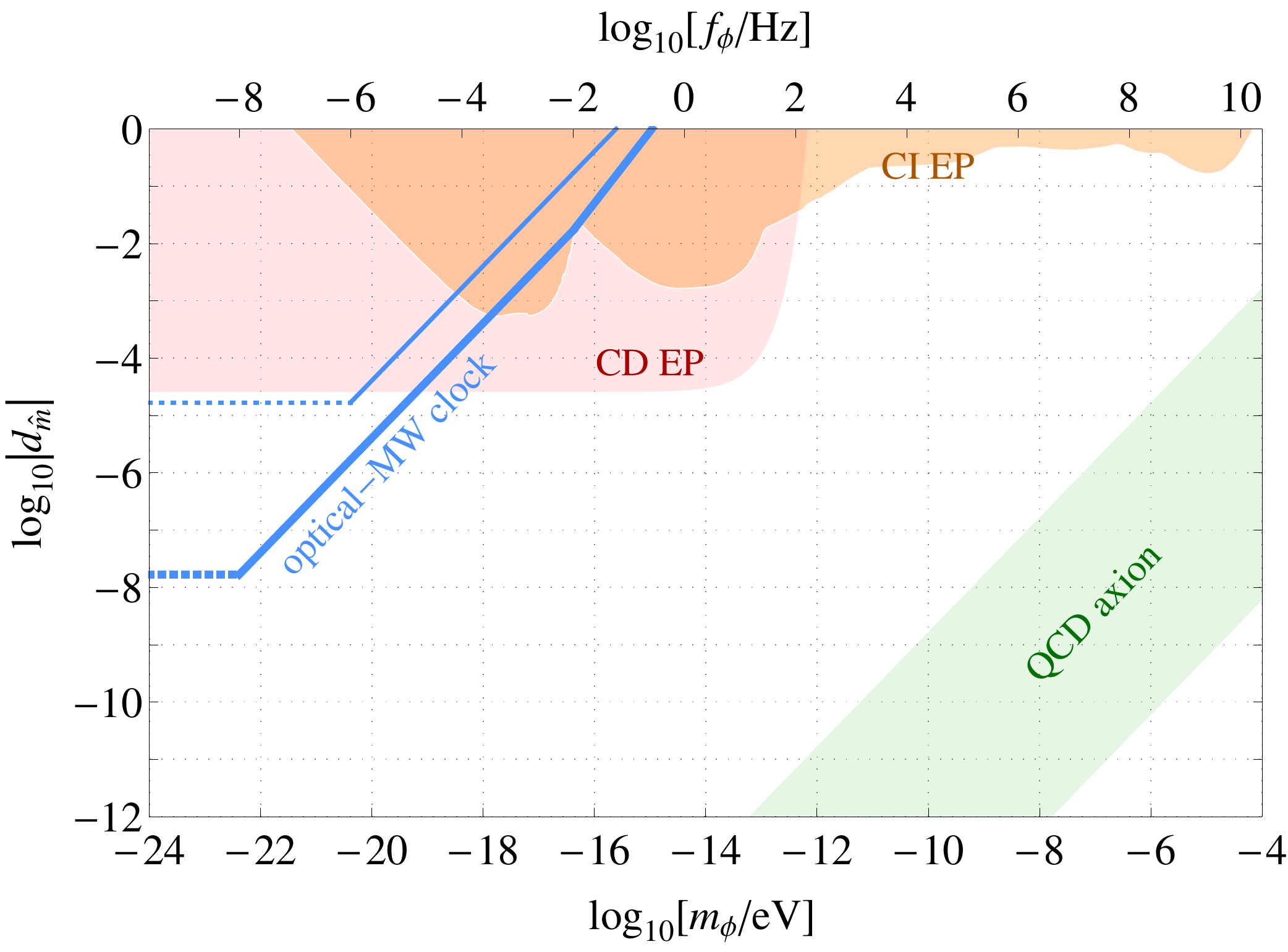}
\caption{(Color online) Sensitivity to $d_{\hat{m}}$ as a function of the scalar dark matter mass $m_\phi = 2\pi f_\phi$ with the microwave-optical clock comparison experiment, assuming $d_{\hat{m}} \gg d_g,d_e,d_{m_e}$. The green region depicts allowed scalar couplings of the QCD axion as in Eq.~\eqref{eq:axion}. Plot labels otherwise similar to Fig.~\ref{fig:dgbound}.}\label{fig:dmqbound}
\end{center}
\end{figure}

\begin{figure}[t]
\begin{center}
\includegraphics[width=0.48\textwidth]{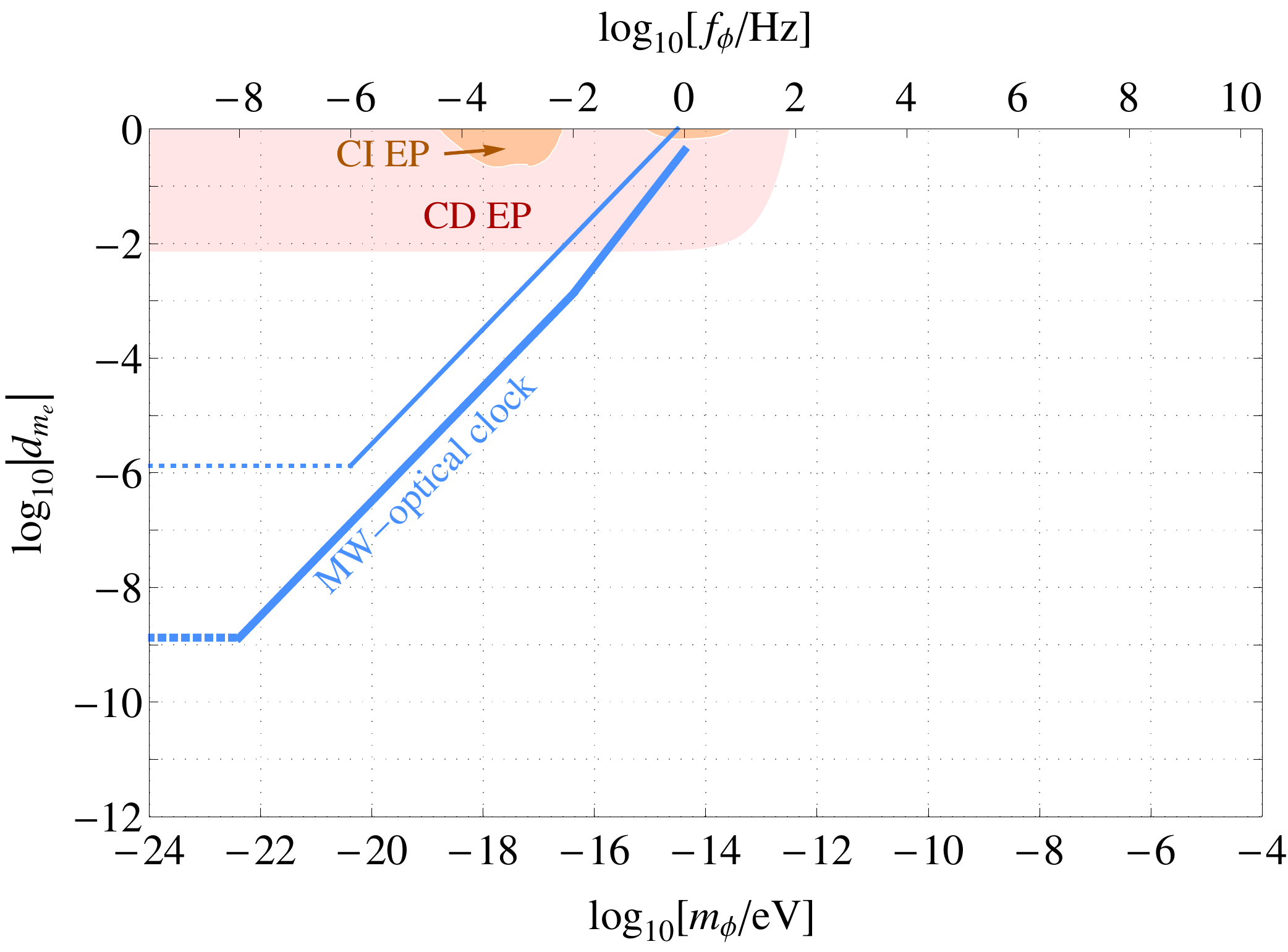}
\caption{(Color online) Sensitivity to $d_{m_e}$ as a function of the scalar dark matter mass $m_\phi = 2\pi f_\phi$ with the microwave-optical clock comparison experiment, assuming $d_{m_e} \gg d_{m_i},d_e,d_g$. Plot labels otherwise similar to Fig.~\ref{fig:dgbound}.
}\label{fig:dmebound}
\end{center}
\end{figure}

For simplicity, we assume in the rest of this subsection that $d_g$, $d_{\hat{m}}$ or $d_{m_e}$ are the only non-zero couplings, so a sensitivity of $\sigma_n \equiv  \sigma_1 / \beta$ on the amplitude of the oscillation in the frequency translates directly into a bound on $d_g F^{1/2}$, $d_{\hat{m}}F^{1/2}$ and $d_{m_e} F^{1/2}$ via Eqs.~\eqref{eq:phiamp}~\&~\eqref{eq:freqratosc}.  We plot the expected $\text{SNR}=1$ sensitivity for the three couplings in Figs.~\ref{fig:dgbound},~\ref{fig:dmqbound}~\&~\ref{fig:dmebound} for integration times $\tau_\text{int}$ of $10^6\,\text{s}$ (thin blue) and $10^8\,\text{s}$ (thick blue), respectively, assuming $\phi$ is all of the dark matter ($F = 1$).  The sensitivity is better for lower masses because the amplitude $\kappa \phi_0$ is bigger in Eq.~\eqref{eq:phiamp}; at high masses, there is an additional suppression due to short coherence times (leading to the ``kink" at $f_\phi \approx 10^{-2} \Hz$ in all three figures)).  
We require $\Delta \tau < \frac{2 \pi}{m_\phi} < \tau_\text{int}$, so the signal shows up as an isolated peak within the range of the DFT of the measurement time series.  Of course, this experiment can still be sensitive to masses for which $\frac{2 \pi}{m_\phi} > \tau_\text{int}$---via a drift in the frequency ratio or a peak in the first mode of the DFT---but any positive signal could not be ascribed unambiguously to the effects of oscillating dark matter. We denoted this by the dotted, flat extension of our sensitivity curves in Figs.~\ref{fig:dgbound},~\ref{fig:dmqbound}~\&~\ref{fig:dmebound}, as the expectation value of the drift over the run time of the experiment is independent of $m_\phi$ in this regime. We note that cosmological constraints (see Sec.~\ref{sec:cosmo}) imply that $\phi$ cannot be all of the DM in the universe for masses $m_\phi \lesssim 10^{-21} \eV$, and that the sensitivity of our proposal scales as $F^{-1/2} = \sqrt{\rho_\text{DM} / \rho_\phi}$ with the fraction $F$ of the local dark matter density in $\phi$.


\subsection{Optical-optical clock comparison}\label{sec:opticaloptical}

\begin{figure}[t] 
\begin{center}
\includegraphics[width=0.48\textwidth]{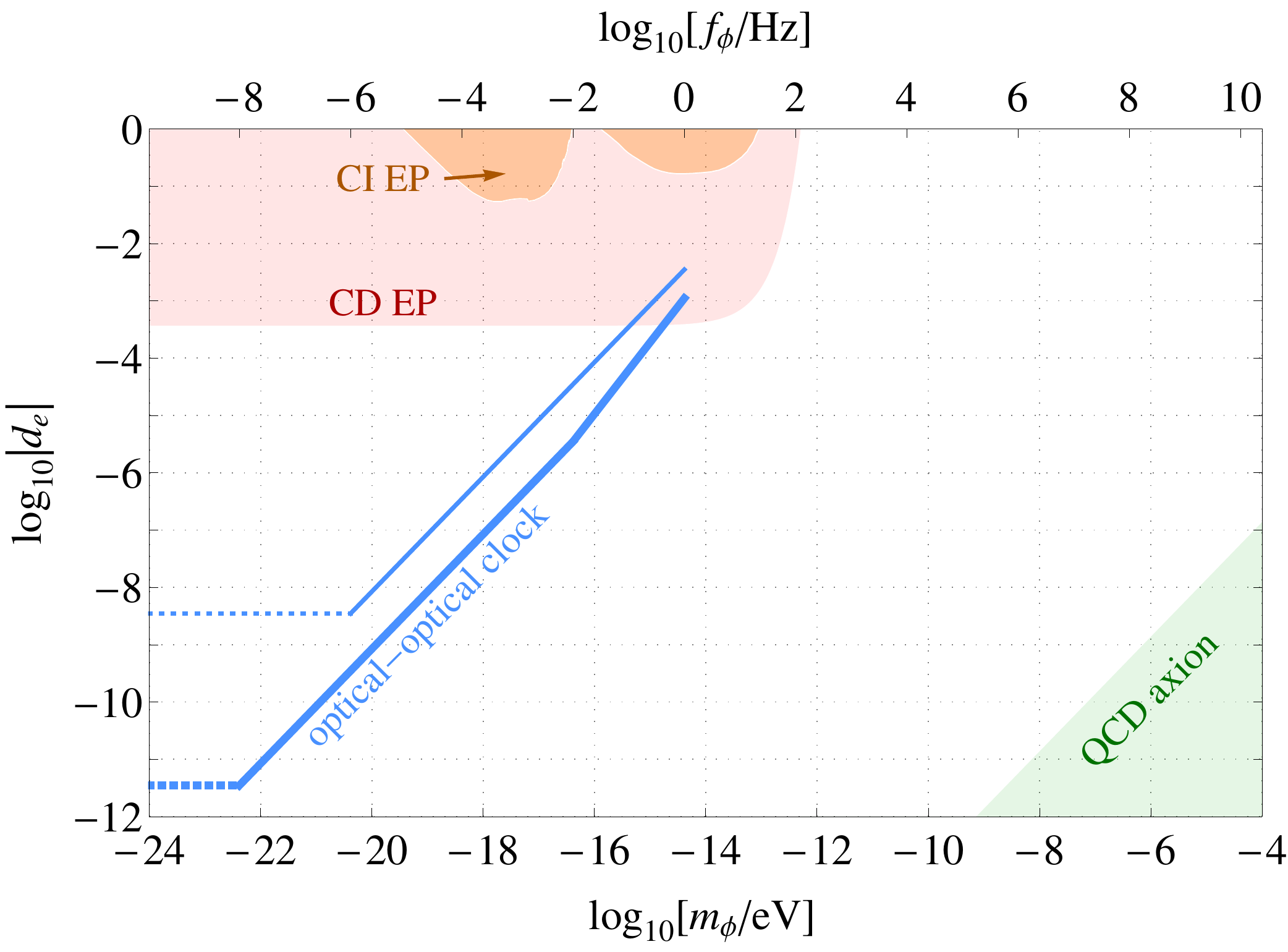}
\caption{(Color online) Sensitivity to $d_e$ as a function of the scalar dark matter mass $m_\phi = 2\pi f_\phi$ with the optical-optical clock comparison experiment described in the text, for $\tau_\text{int} = 10^6,10^8\,s$ (thin blue, thick blue), assuming $\phi$ makes up all of the local dark matter.  Regions excluded by composition-dependent (CD EP) and -independent (CI EP) equivalence principle tests are colored in red and orange, respectively, assuming $d_e \gg d_{m_i},d_g$. The green region depicts allowed scalar couplings of the QCD axion.}\label{fig:debound}
\end{center}
\end{figure}

As explained in Sec.~\ref{sec:setup}, a comparison of two electronic optical transition lines is very sensitive to changes in $\alpha$ alone, and thus the $d_e$ coefficient, without confounding effects due to $d_g$, $d_{\hat{m}}$, or $d_{m_e}$.  We propose measuring variations of the frequency ratio of two optical clocks, one based on a single $^{171}\text{Yb}^+$ ion, the other on a single $^{27}\text{Al}^+$ ion.  Optical clocks based on these ions have demonstrated excellent short-term stability, and behave differently under variations of $\alpha$ because of their dissimilar electron structures and transition dynamics (see Table~\ref{tab:elementlist}).  Specifically, the short-term stability of these clocks can be as good as $2.0 \cdot 10^{-15} \Hz^{-1/2}$ and $2.8 \cdot 10^{-15} \Hz^{-1/2}$ for Yb$^+$ and Al$^+$, respectively, and $\xi_{\text{Yb}^+} - \xi_{\text{Al}^+} \approx -5.3$.  With the same $1\,\Hz$ sampling rate of $f_{\text{Yb}^+} / f_{\text{Al}^+}$, we plot in Fig.~\ref{fig:debound} the $\text{SNR}=1$ sensitivity to the $d_e$ coefficient for integration times of $\tau_\text{int} = 10^6\,\text{s}$ (thin blue) and $\tau_\text{int} = 10^8\,\text{s}$ (thick blue), using the same assumptions as in Sec.~\ref{sec:hyperfineoptical}. Because of the superior short-term stability of optical clocks, the sensitivity to the $d_e$ coupling is strong.  This is particularly exciting given the weaker bounds on $d_e$ from equivalence principle tests, to which we turn in Sec.~\ref{sec:EPtests}. 

\subsection{Improvements in optical clocks}\label{sec:future}

Optical clocks based on \textit{single} ions are nearing their ultimate fractional stability limit due to quantum projection noise (QPN). The path towards further stability improvements thus naturally veers towards using optical clocks based on a large number of \textit{neutral} atoms trapped in optical lattices, whose QPN scales like $N^{-1/2}$ where $N$ is the number of interrogated atoms \cite{PhysRevA.47.3554}.  The most notable examples in this category are optical clocks based on $^{171}$Yb$^0$~\cite{2013arXiv1305.5869H} and $^{87}$Sr$^0$~\cite{Blatt:2008su,Bloom13}, which have demonstrated short-term stabilities of $3 \cdot 10^{-16}\Hz^{-1/2}$ and are approaching long-term stabilities at the $10^{-18}$ level (see Table~\ref{tab:elementlist}). 
Further progress in this area will require pushing the envelope in optical cavity technology for the clock laser, as thermal-noise-induced fluctuations of the cavity length start becoming a limiting factor \cite{jiang2011making,PhysRevLett.109.230801}. Next-generation optical coatings (the dominant thermal noise source) such as microstructured gratings~\cite{PhysRevLett.104.163903} or mirrors based on gallium arsenide~\cite{cole2008apl} may push short-term instabilities below $10^{-17} \Hz^{-1/2}$ \cite{kessler2012sub}.

Optical clock lasers based on a large sample of atoms that do not wholly rely on optical cavities for their short-term stability are also under consideration \cite{hollberg2005optical,hollberg2013}. These systems may reach the QPN-limited instability $ \frac{\Delta \nu}{\nu_0}\frac{1}{ \sqrt{N \tau_1}}$ where $\Delta \nu$ is the spectroscopic linewidth of the clock system, $\nu_0$ is the frequency of the clock transiton, $N$ is the number of atoms measured, and $\tau_1$ is the averaging time in seconds (we have assumed a $ 1\text{\,s}$ measurement time).  With a line quality of ${\Delta \nu}/{\nu_0} \sim 10^{-15}$ and $N \sim 10^{10}$, short-term instabilities of $\sim 10^{-20} \Hz^{-1/2}$ may be within reach.\footnote{Private communication with Leo Hollberg.}  Furthermore, advanced systems using arrays of coherent atomic samples may exhibit Heisenberg-limited performance, for which the instability $ \frac{\Delta \nu}{\nu_0}\frac{1}{ N \sqrt{\tau_1}}$ could be as low as $10^{-21} \Hz^{-1/2}$ with only $N\sim 10^6$ atoms \cite{hollberg2005optical}.

The difference $\xi_\text{A} - \xi_\text{B}$ in the coefficients of Eq.~\eqref{eq:coeffpowers} is typically small if A and B are both transitions in neutral atoms, as seen in Table~\ref{tab:elementlist}, leading to a reduced sensitivity to $d_e$ relative to comparisons between two ion clocks (or an ion-atom clock system) for the same instability.  However, at their current rate of stability improvements,  optical clocks based on neutral atoms will likely lead to better potential sensitivity to $d_e$ in the future. 


\subsection{Nuclear clocks} \label{sec:nuclear}

It has been suggested in \cite{PhysRevLett.108.120802,2003EL.....61..181P} that a \textit{nuclear} clock based on a narrow isomer transition in $\mathrm{{}^{229\text{m}}Th}$ may be used to set a better bound on drifts of fundamental constants.  The thorium nucleus has the remarkable property of having an excited isomer state of only $7.6 \pm 0.5 \eV$ and linewidth of $\sim 10^4 \Hz$, accessible to current lasers.\footnote{It must be noted that the $7.6 \eV$ thorium line has not yet been directly observed; the size of the gap has been determined via indirect measurements of nuclear decays of uranium \cite{PhysRevLett.98.142501}.} 
The small gap of $7.6\eV$ between the isomer and ground state---typically $\mathcal{O}(100\keV)$ for most nuclei---arises due to an accidental cancellation between contributions from electromagnetic and strong interactions \cite{PhysRevLett.97.092502,Flambaum:2008ij,Berengut:2009zz,Wense:2012ey,PhysRevLett.109.160801}. This leads to an enhancement in sensitivity to changes in $\alpha$, $\Lambda_3$, and quark masses \cite{1742-6596-264-1-012010}:
\begin{equation}\label{eq:thoriumsens}
\frac{\delta {f_\text{Th} }}{f_\text{Th}} \approx 10^4 \left(d_e + 10 (d_g - d_{\hat{m}})+ \dots \right),
\end{equation}
where $d_{\hat{m}}$ is the dilaton coupling to the symmetric combination of the quark masses as in Eq.~\eqref{eq:dmhat}.

\begin{figure}[t] 
\begin{center}
\includegraphics[width=0.48\textwidth]{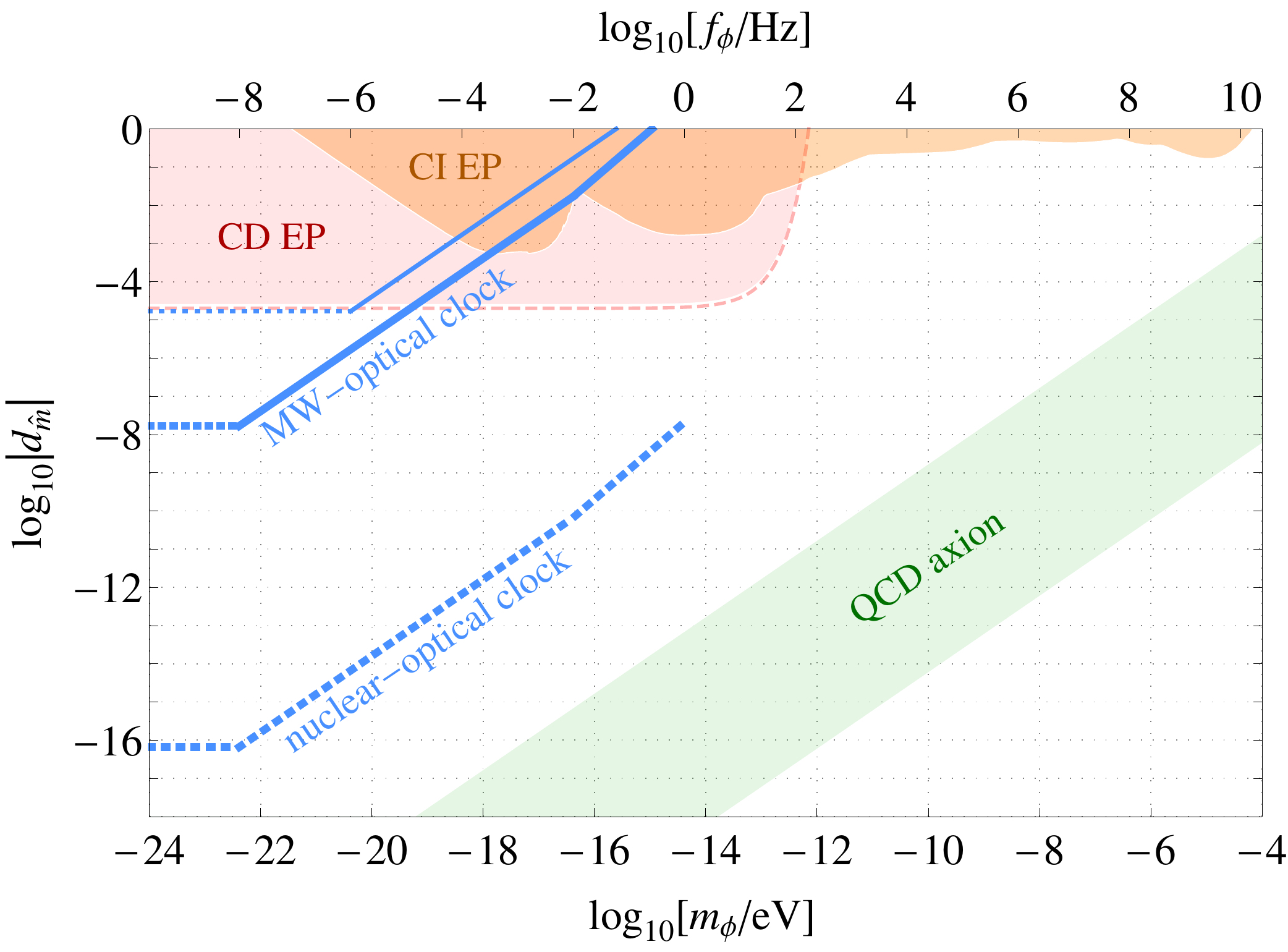}
\caption{(Color online) Same as in Fig.~\ref{fig:dmqbound}, adding the reach on $d_{\hat{m}}$ with a future nuclear-optical clock comparison after $\tau_\text{int} = 10^{8}\text{\,s}$. The future $d_{\hat{m}}$ sensitivity of the composition-dependent EP test in \cite{Dimopoulos:2006nk} is shown as a dashed red line.  }\label{fig:dmqboundfut}
\end{center}
\end{figure}

\begin{figure}[t] 
\begin{center}
\includegraphics[width=0.48\textwidth]{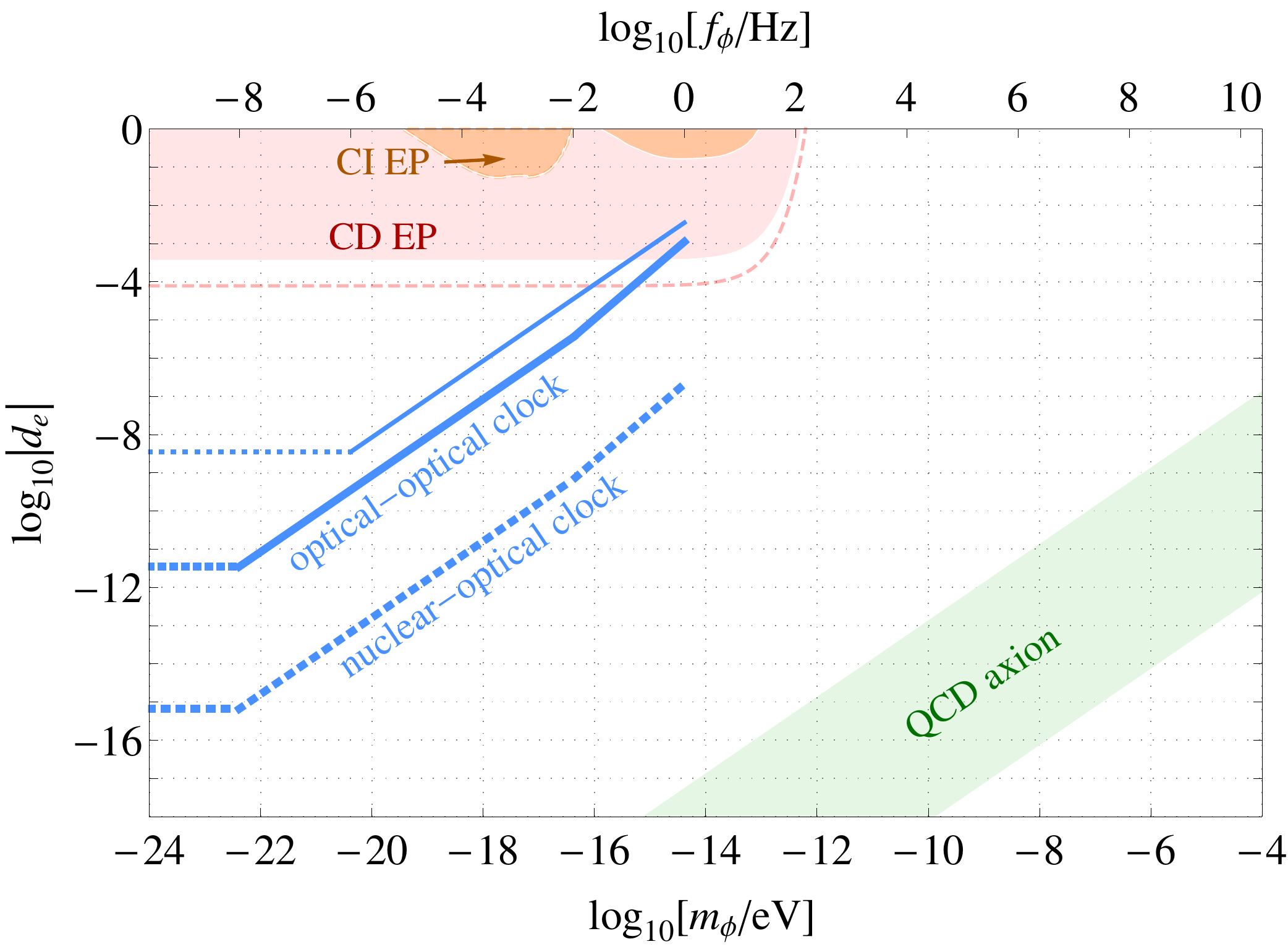}
\caption{(Color online) Same as in Fig.~\ref{fig:debound}, adding the reach on $d_e$ with a future nuclear-optical clock comparison after $\tau_\text{int} = 10^{8}\text{\,s}$.  The future $d_e$ sensitivity of the composition-dependent EP test in \cite{Dimopoulos:2006nk} is shown as a dashed red line.}\label{fig:deboundfut}
\end{center}
\end{figure}

Proposals have been put forward to build a solid-state thorium nuclear clock, using either $\mathrm{{}^{229}Th^{2+}}$ ions doped inside a $\mathrm{CaF_2}$ lattice~\cite{2012NJPh...14h3019K}, or a single-ion clock based on $\mathrm{{}^{229}Th^{3+}}$ \cite{PhysRevLett.108.120802}. The latter proposal could reach its quantum-limited stability of $10^{-15} \Hz^{-1/2}$ for interrogation times of a second or longer, if thermal noise can be controlled.  Comparison with a line from an optical clock is feasible given the recent development of ``vacuum ultraviolet" frequency combs \cite{PhysRevLett.100.253901,2012Natur.482...68C}.  Anticipating these technologies to mature in the next decade, we project a combined uncertainty of $\sim 10^{-15}\Hz^{-1/2}$ for the optical synthesis process and stability of both the nuclear and optical clock.  Translating this to sensitivity to the couplings in Eq.~\eqref{eq:couplings}, we project minimal reach with a nuclear-optical clock comparison (in dotted blue) for $d_{\hat{m}}$, $d_e$ and $d_g$ and  in Figs.~\ref{fig:dmqboundfut},~\ref{fig:deboundfut}~\&~\ref{fig:dgboundfut}, respectively.

\section{Equivalence principle tests} \label{sec:EPtests}

Ultralight scalars can mediate long-range Yukawa forces between uncharged objects, and will thus result in deviations from the equivalence principle (EP).  In the parametrization of \cite{Damour:2010rp}, a scalar (in addition to gravity) will create the potential
\begin{align} \label{eq:eppotential}
V  &= -G \frac{m_A m_B}{r_{AB}} \left(1 + \alpha_A \alpha_B e^{-m_\phi r_{AB}}\right); \nonumber \\
\alpha_A &= \pd{ \ln \left[\kappa m_A(\kappa \phi)\right] }{\kappa \phi} \equiv d_g + \overline{\alpha}_A.
\end{align}

For macroscopic objects, most of the rest mass comes from the nucleus mass, so unless $d_g$ is suppressed relative to the $d_{m_i}$ or $d_e$, we expect $\alpha_A \simeq d_g \gg \overline{\alpha}_A$.  In this case, $\alpha_A$ is not strongly dependent on the chemical composition of object $A$, but one can still detect a scalar force by looking for a departure from $V \propto r^{-1}$ in Eq.~\eqref{eq:eppotential} in experiments with linear size of order $m_{\phi}$. For a review, we refer the reader to \cite{Fischbach:1996eq}, from which we displayed a compendium of composition-independent bounds on $d_g$ as the orange region in Fig.~\ref{fig:dgbound}. Composition-independent tests of the EP can also directly constrain $d_{\hat{m}}$, $d_{m_e}$ and $d_e$.  Without any extra assumptions, these constraints are weaker because of the relatively small quark-mass, electron-mass, and electromagnetic contributions to the rest mass of atoms.  We can quantify this by writing
\begin{align}\label{eq:alphaCD}
\overline{\alpha}_A \equiv \Big[&+(d_{\hat{m}} - d_g) Q_{\hat{m}} +  (d_{\delta {m}} - d_g) Q_{\delta{m}} \nonumber \\
&+ (d_{m_e} - d_g) Q_{m_e} + d_e Q_e\Big]_A,
\end{align} 
where $d_{\hat{m}} = \frac{d_{m_d} m_d +d_{m_u} m_u}{m_d + m_u}$ and $d_{\delta {m}} = \frac{d_{m_d} m_d - d_{m_u} m_u}{m_d - m_u}$ are the dilaton couplings to the symmetric and antisymmetric combination of the quark masses respectively. The ``charge vector" $Q_{\hat{m}} \equiv \pd{\ln m_A}{\ln \hat{m}}$ is $\sim 0.1$ and decreases by $\sim 10^{-2}$ for elements with high atomic number $A$, while $Q_{\delta{m}}$ is small and nearly constant across the periodic table. The charge vector $Q_{m_e} \equiv \pd{\ln m_A}{\ln m_e}$ is also nearly constant at the rough value of $ 2.5 \cdot 10^{-4}$, while $Q_{e} \equiv \pd{\ln m_A}{\ln \alpha}$ ranges from $3\cdot 10^{-4}$ for the lightest elements to $4\cdot 10^{-3}$ for heavy elements ($Z\gtrsim50$) (see \cite{Damour:2010rp} for formulas of the $Q_i$). We therefore estimate that the minimal composition-independent EP bounds on $d_{\hat{m}}$, $d_{m_e}$ and $d_e$ are weaker than those on $d_g$ by factors of $10$, $4 \cdot 10^{3}$ and $10^{3}$, respectively. We indicated them as orange regions in Figs.~\ref{fig:dmqbound},~\ref{fig:dmebound}~\&~\ref{fig:debound}. 

At distances larger than the Earth's radius, \textit{composition-dependent} EP tests are more constraining than the composition-independent ones.  Most notably, Lunar Laser Ranging (LLR) \cite{Williams:2004qba,Williams:2012nc}, which measures the differential acceleration of the Earth and Moon in the Sun's gravitational field, and the E\"otWash experiment \cite{Schlamminger:2007ht,Wagner:2012ui}, which measures the differential acceleration for Be and Ti on Earth, are both sensitive to fractional differential accelerations of order $10^{-13}$. Because LLR is done at a larger length scale and the Earth and Moon have similar chemical composition, the E\"{o}tWash experiment is typically more constraining, setting a limit $\left|\alpha_{\text{Earth}} (\alpha_\text{Be} -\alpha_\text{Ti} ) \right| \lesssim 3.6 \cdot 10^{-13}$ at 95\% CL. For $d_g \gtrsim d_{m_i},d_e$, this constrains $|d_g| \lesssim 7.2 \cdot 10^{-6}$ for $m_{\phi} \lesssim 1/R_\text{Earth}$ \cite{Damour:2010rp}. Similarly, with $d_{\hat{m}}$, $d_{m_e}$ or $d_e$ the only non-zero couplings in Eq.~\eqref{eq:alphaCD}, minimal bounds of $|d_{\hat{m}}| \lesssim 2.5 \cdot 10^{-5}$, $|d_{m_e}| \lesssim 7.1 \cdot 10^{-3}$ and $|d_e| \lesssim 3.6 \cdot 10^{-4}$ are obtained. 
The atom-interferometric experiment proposed in \cite{Dimopoulos:2006nk} (currently under construction) will improve the minimal coupling bounds to $|d_{\hat{m}}| \lesssim 2.0 \cdot 10^{-5}$, $|d_{m_e}| \lesssim 8.5 \cdot 10^{-4}$ and $|d_e| \lesssim 7.8 \cdot 10^{-5}$ once it reaches initial design precision of $ \left|\alpha_{\text{Earth}}(\alpha_{^{85}\text{Rb}} -\alpha_{^{87}\text{Rb}}) \right| \lesssim 10^{-15}$. The minimal composition-dependent EP constraints are depicted as red regions in Figs.~\ref{fig:dgbound},~\ref{fig:dmebound}~\&~\ref{fig:debound}.  The future reach on $d_{\hat{m}}$ and $d_e$ with the proposal of \cite{Dimopoulos:2006nk} is shown as a thin dashed red line in Figs.~\ref{fig:dmqboundfut}~\&~\ref{fig:deboundfut}.

\section{Gravitational antennas}\label{sec:gravwaves}

There are two ways in which the scalar $\phi$ of Eq.~\eqref{eq:couplings} can influence the macroscopic motions of matter.\footnote{In \cite{Rubakov2013}, it is suggested that future millisecond pulsar timing measurements may become sensitive to the oscillating pressure created by a scalar dark matter field (cfr.~the $\gamma_p$ term in Sec.~\ref{sec:production}) in a narrow mass range around $\sim 10^{-23}\eV$, purely from the gravitational coupling of $\phi$.} 
 We discussed the first in Sec.~\ref{sec:EPtests}: apparent violations of the equivalence principle through new scalar forces, which do \textit{not} hinge on the assumption that $\phi$ makes up the dark matter density. \textit{With} this assumption, however, spatial gradients in the non-relativistic $\phi$ waves can cause tidal forces not unlike those produced by gravitational waves (GW).  In what follows, we will quantify the sensitivity of GW detectors such as LIGO \cite{Abbott:2007kv,LigoS6}, Advanced LIGO \cite{0900288,Harry:2010zz}, AGIS \cite{Dimopoulos:2008sv,2011GReGr..43.1953H}, and eLISA \cite{AmaroSeoane:2012je} to ultralight scalar dark matter waves.

To leading order, the potential for an otherwise free-falling test mass $M$ (e.g.~a LIGO mirror) in a scalar wave $\phi(t, \vec{x}) = \phi_0 \cos(m_\phi t - \vec{k_\phi} \cdot \vec{x} + \alpha)$ is
\begin{align} \label{eq:scalarpotential}
V = M\left[ 1 + \alpha_M \kappa \phi(t,\vec{x}) \right] \simeq M \left[ 1 + d_g \kappa \phi(t,\vec{x}) \right],
\end{align}
with $\kappa \equiv \frac{\sqrt{4\pi}}{\MPl}$, and where we have approximated $\alpha_M \simeq d_g$ (cf.~Eq.~\eqref{eq:eppotential}) since we care about test masses composed of neutral atoms, whose rest mass is primarily determined by the QCD scale $\Lambda_3$.  For simplicity, let us assume the scalar wave travels in the $x$-direction, $\vec{k_\phi} = k_\phi \hat{x}$.  Then the potential in Eq.~\eqref{eq:scalarpotential} will cause test masses to deviate from their geodesics $\vec{x}_M$ by a displacement:
\begin{align}
\delta x_M \sim \frac{d_g k_\phi \kappa \phi_0}{m_\phi^2} \sin(m_\phi t - k_\phi x_M + \alpha).
\end{align}

Between two mirrors positioned at $x_M \simeq 0$ and $x_M \simeq L$, the return trip time for light between the two mirrors is approximately $t_\text{return} \simeq 2L + \delta t$ with:
\begin{align}
\delta t \simeq \frac{4 d_g k_\phi \kappa \phi_0}{m_\phi^2}  \sin^2\left( \frac{m_\phi L}{2} \right) \sin(m_\phi L + \alpha),
\end{align}
where we have taken $v \ll \min\lbrace 1, m_\phi L \rbrace$.

\begin{figure}[t] 
\begin{center}
\includegraphics[width=0.48\textwidth]{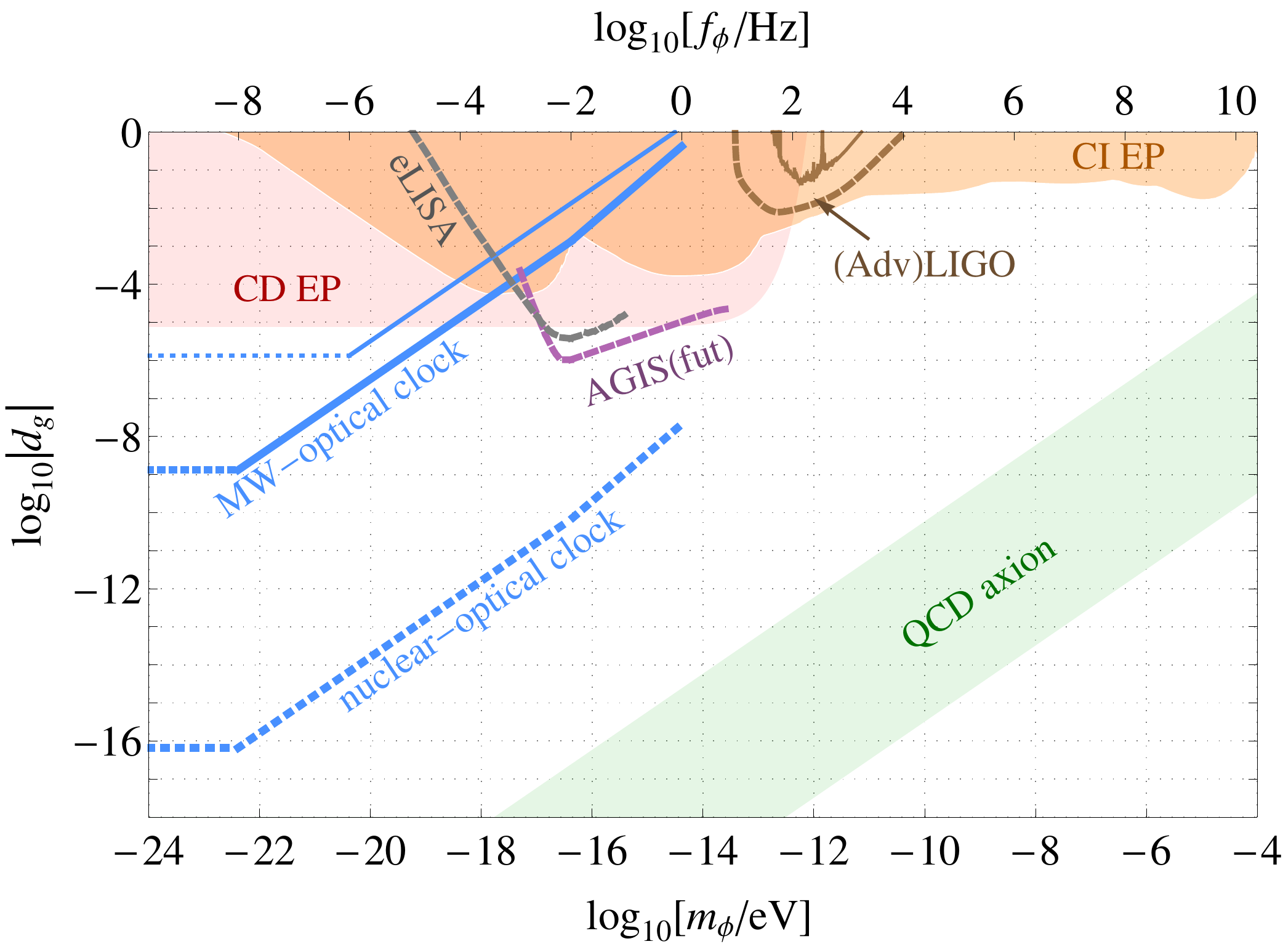}
\caption{(Color online) Same as in Fig.~\ref{fig:dgbound}, adding the reach on $d_g$ with a future nuclear-optical clock comparison after $\tau_\text{int} = 10^{8}\text{\,s}$.  The estimated sensitivity to $d_g$ with current and proposed gravitational wave detectors is also depicted: LIGO (brown), Advanced Ligo (dashed brown), AGIS-Future (dashed purple), eLISA (dashed gray). }\label{fig:dgboundfut}
\end{center}
\end{figure}

A gravitational wave with $\oplus$-polarization moving in the $z$-direction with phase $\alpha$, strain amplitude $h_0$, and angular frequency $\omega$, causes an analogous time delay of $\delta t = \frac{h_0}{\omega} \sin(\omega L) \sin(\omega L + \alpha)$.  Hence, we find that (up to angular factors), a scalar wave has an effective GW strain amplitude
\begin{align} \label{eq:effstrain}
h_{0,\text{SW}} \sim \frac{2 d_g k_\phi \kappa \phi_0}{m_\phi}  \tan\left( \frac{m_\phi L }{2} \right),
\end{align}
and angular frequency $\omega = m_\phi$. Of course, the angular response function of a GW detector is different for a scalar wave. After averaging over the appropriate antenna patterns and taking into account that the scalar wave is expected to have a wave vector $\vec{k_\phi} \simeq m_\phi \vec{v}$ and coherence time $\tau_\text{coh} \simeq 2\pi (m_\phi v^2)^{-1}$ with $v \approx 10^{-3}$, we estimated the sensitivity of various interferometors to scalar dark matter waves with $d_g$ couplings using Eq.~\eqref{eq:effstrain}. (The sensitivity to $d_{m_e}$ and $d_e$ is much weaker because the charge vectors $Q_{m_e}$ and $Q_e$ defined in Sec.~\ref{sec:EPtests} are small.)  The resulting estimated reach is plotted in Fig.~\ref{fig:dgboundfut}.  In general, we find that larger interferometers such as the large version of AGIS \cite{Dimopoulos:2008sv} and eLISA \cite{AmaroSeoane:2012je} have better sensitivity, because the effective strain amplitude in Eq.~\eqref{eq:effstrain} increases linearly with interferometer size $L$ (assuming $L \ll m_\phi^{-1}$).  For this reason, smaller detectors such as (Advanced) LIGO and AGIS-LEO \cite{2011GReGr..43.1953H} cannot surpass current EP experiments in terms of sensitivity to our model (unless there is a local overdensity of $\phi$ dark matter).\\

\section{Cosmology and astrophysics}\label{sec:cosmo}

\subsection{Cosmic production and evolution}\label{sec:production}

The energy associated with coherent oscillations of light scalars can play the role of the dark matter energy density.  Light scalar dark matter can be non-thermally produced through vacuum misalignment in the early universe (``misalignment mechanism") \cite{Linde:1987bx}. The evolution of a light scalar field with initial amplitude $\phi_{0,i}$ can be described classically (because of high occupation numbers) by its equation of motion:
\begin{align}\label{eq:KG}
\ddot{\phi} + 3 H \dot{\phi} + V'(\phi) = 0,
\end{align}
where the Hubble scale is defined as $H = \frac{\dot{a}}{a}$, and $V(\phi) = \frac{1}{2}m_\phi^2 \phi^2 + \frac{a_\phi}{3} \phi^3 + \frac{\lambda_\phi}{4} \phi^4$ as before. When the Hubble scale is very large, specifically when $H^2 \gg V'(\phi_{0,i})/\phi_{0,i}$, the scalar field and its energy density will be ``frozen" at $\phi_{0,i}$ and $V(\phi_{0,i})$, respectively.  In this regime, $V(\phi_{0,i})$ acts as a contribution to the vacuum energy. As the universe expands and the Hubble scale drops well below the critical value ($H^2 \simeq V'(\phi_{0,i})/\phi_{0,i}$), the field will start oscillating with negligible friction. If the oscillations are harmonic---when the $m_\phi^2 \phi^2 / 2$ term dominates---the energy density of oscillations  $\rho_\phi \simeq m_\phi^2 \phi^2_0 /2 $ with amplitude $\phi_0$ acts as cold, pressureless dark matter on length scales larger than the field's Compton wavelength $\sim 2\pi / m_\phi$.

The one-loop effective potential $V(\phi)$ has to satisfy several constraints if $\phi$ is to be a viable dark matter candidate. 
Most obviously, we want to avoid runaway behavior of $\phi$ or convergence to minima other than $\phi = 0$ in our parametrization of $V(\phi)$. It is likely sufficient to require anharmonicities to be small when the scalar starts oscillating at $H \sim m_\phi$, i.e.~$|a_{\phi} \phi_{0,i}|, |\lambda_\phi \phi_{0,i}^2| < m_{\phi}^2$, although this is not a strict condition, as unknown Planckian dynamics and higher-dimensional operators may be important at those high field values.

The couplings to the SM in Eq.~\eqref{eq:couplings} radiatively generate self-interaction parameters of order
\begin{align}
\delta a_{\phi, \mathrm{SM}} &\sim  - \frac{ \left( d_{m_i} \kappa m_{i} \right)^3 m_i}{16\pi^2} \sim - 10^{-58} \eV \left(\frac{ d_{m_t}}{10^{-6}}\right)^3 \label{eq:selfrad1} \\
\delta \lambda_{\phi, \mathrm{SM}} &\sim + \frac{\left( d_{m_i} \kappa m_{i} \right)^4}{16\pi^2}  \sim + 10^{-92} \left(\frac{ d_{m_t}}{10^{-6}}\right)^4 \label{eq:selfrad2}
\end{align}
where,  to get numerical estimates,  we have plugged in the top quark mass $m_t$ and corresponding coupling $d_{m_t}$---probably the biggest SM correction (the Higgs boson contributes with opposite sign). In addition, gravitational effects will introduce an effective quartic of $ \delta \lambda_{\phi,\text{grav}} \sim - m_\phi^2/\MPl^2 \approx -10^{-92}  \left(\frac{m_{\phi}}{10^{-18} \eV}\right)^{2}$.

For the remainder of the discussion, we will combine the couplings $a_\phi$ and $\lambda_\phi$ into an effective quartic coupling
\begin{widetext}
\begin{align}
\lambda_\phi^{\mathrm{eff}} &\equiv \delta \lambda_{\phi,\mathrm{SM}} - \frac{10 \delta a_{\phi,\mathrm{SM}}^2}{9 m_\phi^2} + \delta \lambda_{\phi,\text{grav}} \nonumber \\
&\sim + 10^{-92} \left(\frac{ d_{m_t}}{10^{-6}}\right)^4 - 10^{-80} \left(\frac{ d_{m_t}}{10^{-6}}\right)^6   \left(\frac{10^{-18} \eV}{m_{\phi}}\right)^{2} - 10^{-92}  \left(\frac{m_{\phi}}{10^{-18} \eV}\right)^{2},
\label{eq:selfeff}
\end{align}
\end{widetext}
serving as an effective measure of attractive ($\lambda_\phi^{\text{eff}} < 0$) or repulsive ($\lambda_\phi^{\text{eff}} > 0$) self-interactions of $\phi$ in the non-relativistic limit.
The cubic contribution (which is always negative) in Eq.~\eqref{eq:selfeff} tends to dominate for the parameter range of interest in this paper, implying that $\phi$ typically has attractive self-interactions. Imposing that the potential be harmonic when the field $\phi$ enters its oscillating phase at $H \sim m_\phi$ (anharmonicities today will be much smaller because the density redshifts as $\rho_\phi \propto a^{-3}$) to avoid runaway behavior yields the condition:
\begin{align}
\left|\lambda_\phi^{\mathrm{eff}}\right| \lesssim \frac{m_\phi^2}{\phi_{0,i}^2} \lesssim 10^{-86} \left(\frac{\rho_\text{DM,U}}{\rho_\phi} \right) \left(\frac{m_\phi}{10^{-18} \eV }\right)^{5/2}.
\label{eq:quareff}
\end{align}
We note that this is a weak condition, as higher-dimensional operators may stabilize the full one-loop effective potential at high field values.  However, it is striking that couplings $d_i \lesssim 10^{-6}$ in Eq.~\eqref{eq:quareff}, within reach of our proposed clock comparison experiment, are roughly in the right ballpark to satisfy the inequality \ref{eq:quareff} if $\phi$ comprises all of the DM in the universe $\rho_\phi = \rho_\text{DM,U}$.

\subsection{Structure formation}\label{sec:structure}

The field $\phi$ should act as dark matter today, so it must presently be in its oscillating phase, implying that $m_\phi \gtrsim H \approx 10^{-33} \eV$.  If the transition to the oscillating regime happens close to the surface of last scattering, the integrated Sachs-Wolfe effect will cause distortions in the anisotropy spectrum of the cosmic microwave background \cite{Ferreira:1997hj}. This phenomenon provides a handle to exclude light scalars as a significant component of dark matter for masses below $10^{-26} \eV$ \cite{Amendola:2005ad}. 

The most stringent cosmology limits on such ultralight scalars are due to the fact that they are not entirely pressureless dark matter \cite{Turner:1983he}, because of the tiny mass and  self-interactions of $\phi$. In the oscillating phase, the pressure $p_\phi \equiv \dot{\phi}^2/2 - V(\phi)$ will vary much on time scales shorter than $H^{-1}$ and can have a non-zero average over one oscillation cycle of the field $\phi$ when the potential is anharmonic $a_\phi,\lambda_\phi \neq 0$. Following \cite{Turner:1983he}, we quantify this as $p_\phi = (\gamma+\gamma_p - 1 ) \rho_\phi$, where $\gamma$ is the average of $(p_\phi+\rho_\phi)/\rho_\phi = \dot{\phi}^2/V(\phi_0)$ over one oscillation cycle, and $\gamma_p$ indicates the oscillatory piece, irrelevant on large enough time and length scales.  Treating the anharmonic terms as small corrections, one finds:
\begin{widetext}
\begin{align}
\gamma &=\frac{1}{T} \int_0^T \frac{\dot{\phi}^2}{V(\phi)} dt \simeq 2 \frac{ \int_{-\phi_0}^{+\phi_0} \left[1-\frac{V(\phi)}{V(\phi_0)}\right]^{+1/2} d\phi}{ \int_{-\phi_0}^{+\phi_0} \left[1-\frac{V(\phi)}{V(\phi_0)}\right]^{-1/2} d\phi } \approx 1 + \frac{3 \lambda_\phi^{\mathrm{eff}} \phi_0^2}{16 m_\phi^2} + \mathcal{O}\left(\frac{{\lambda_\phi^{\mathrm{eff}}}^2 \phi_0^4}{m_\phi^4}\right)
\end{align}
\end{widetext}
In the next subsection, we shall discuss the consequences of the resulting average pressure on large-scale structure formation.

Density perturbations of ultralight scalar fields have non-zero sound speeds at short length scales.  There is a well-known scale-dependent sound speed $c_{s,(m)}^2 = \text{max}\left\lbrace 1, | {\bf{k}} |^2 / (4 m_\phi^2 a^2)\right\rbrace $ \cite{Hu:2000ke}, as well as a scale-independent sound speed contribution $c_{s,(\lambda)}^2 = \gamma - 1$ when the average pressure is non-zero \cite{Turner:1983he}.  Density perturbations $\delta_{\bf{k}} \equiv \delta \rho_{\bf{k}} / \rho $ at scales $|\bf{k}|$ obey the approximate classical evolution equation:
\begin{align}\label{eq:growth}
\ddot{\delta}_{\bf{k}} + 2 H \dot{\delta}_{\bf{k}} \simeq \left[\frac{4 \pi \rho_\text{tot}}{\MPl^2} - \left(  \frac{\bf{k}^2}{4 m_\phi^2 a^2} +\frac{3 \lambda_\phi^{\mathrm{eff}} \phi_0^2}{16m_\phi^2} \right) \frac{\bf{k}^2}{a^2} \right] \delta_{\bf{k}}.
\end{align}
At large length scales (small $|\bf{k}|$), this evolution equation describes the growth of structure $\delta_{\bf{k}} \propto a$ in the presence of cold dark matter after matter-radiation equality. At short length scales, however, Eq.~\eqref{eq:growth} does no longer admit linear growth $\delta_{\bf{k}}$ solutions.  The $c_{s,(m)}^2$ term will inhibit growth at length scales below a Jeans length $L_{J,(m)}$, while the $c_{s,(\lambda)}^2$ term can also inhibit ($\lambda_\phi^{\mathrm{eff}} > 0$) or exponentiate ($\lambda_\phi^{\mathrm{eff}} < 0$) growth at scales shorter than $L_{J,(\lambda)}$, where
\begin{align}
L_{J,(m)} \simeq \left(\frac{\pi^3 \MPl^2}{\rho_\phi m_\phi^2} \right)^{1/4}; ~~ L_{J,(\lambda)} \simeq \left( \frac{3 \pi |\lambda_\phi^{\mathrm{eff}}| \MPl^2}{8 m_\phi^4} \right)^{1/2},
\end{align}
and we have assumed the scalar $\phi$ makes up all of the dark matter density  ($\rho_\phi \simeq \rho_\text{tot}$).  As a conservative estimate, we posit that growth on length scales larger than current galactic sizes could not have been disturbed by the above growth inhibition (or acceleration).  
Specifically, we require $L(a) >  \max \lbrace L_{J,(m)}(a), L_{J,(\lambda)}(a) \rbrace$ for any length scale $L(a_0) \gtrsim 100 \, \text{kpc}$ at any time between matter-radiation equality and the present ($a_\text{eq} < a < a_0$). Note that in the matter dominated era, we have the scalings $L_{J,(m)}(a) \propto a^{3/4}$, $L_{J,(\lambda)}(a) \propto a^0$, so the above condition is strongest at $a = a_\text{eq}$.  We thus obtain estimated bounds on the potential parameters:
\begin{align} \label{eq:structurelimit}
m_\phi \gtrsim 6 \cdot 10^{-21} \eV; ~~ |\lambda_\phi^{\mathrm{eff}}| \lesssim 3 \cdot 10^{-79} \left(\frac{m_\phi}{10^{-18} \eV }\right)^4.
\end{align}
We note that these constraints disappear completely when $\phi$ is only a sub-dominant component of dark matter, and that these estimates are more conservative than other recent studies.

\begin{figure}[t]
\begin{center}
\includegraphics[width=0.48\textwidth]{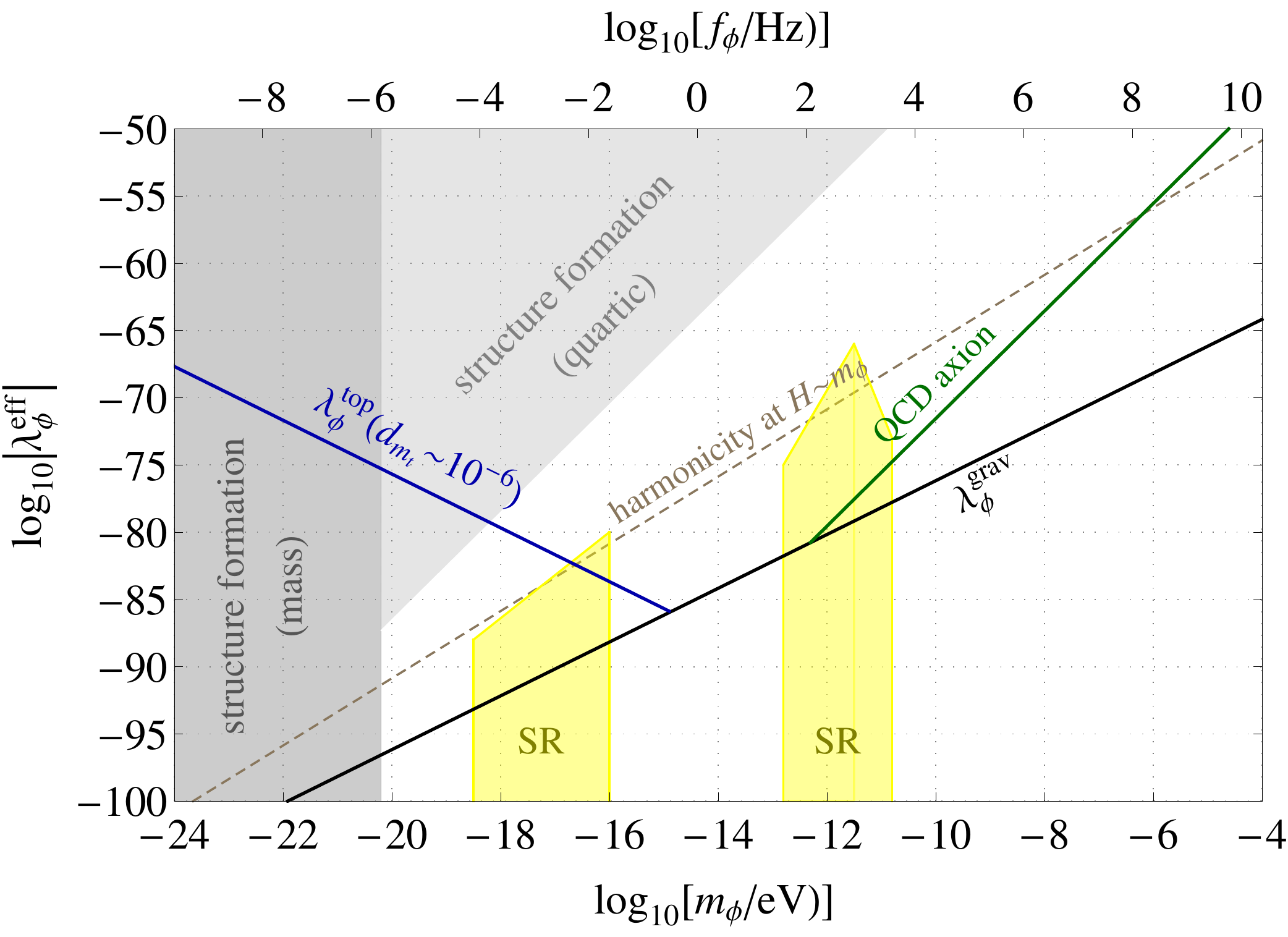}
\caption{(Color online) Cosmological and astrophysical constraints in the $m_\phi$--$|\lambda_\phi^{\text{eff}}|$ plane.  Structure formation constraints are indicated in dark and light gray.  Superradiance (SR) bounds from observations of supermassive black holes are depicted as the two left most yellow regions, while the bounds from observations of stellar mass black holes are depicted as the right most yellow region. 
For reference, we include the effective quartic contribution generated for $\phi$ when $d_{m_t} \sim 10^{-6}$ from SM corrections (blue), the gravitational contribution to the effective quartic for a massive scalar field (black), as well as the mass-quartic relation for the QCD axion (green). The structure formation bounds assume $\rho_\phi \sim \rho_\text{DM}$, and disappear when $\rho_\phi \lesssim 10^{-1} \rho_\text{DM}$.}\label{fig:cosmobound}
\end{center}
\end{figure}

Observations of the Lyman $\alpha$ flux power spectrum with the Keck High Resolution Echelle Spectrometer (HIRES) and the Magellan Inamori Kyocera Echelle (MIKE) imply the existence of hydrogen clouds with a comoving size down to around $100\,\text{kpc}$ at high redshift~\cite{Viel:2013fqw}, and were used in \cite{Marsh:2013ywa} to exclude scalar fields lighter than $\sim 10^{-21} \eV$ as the dominant component of dark matter,\footnote{The limit in \cite{Marsh:2013ywa} relies on an analogy between scalar and warm dark matter, and only gives an order of magnitude correct exclusion limit; detailed hydrodynamical simulations of structure formation with light scalar DM are needed to make this limit more precise \cite{Marsh:2011bf}.}
consistent with our estimate in Eq.~\eqref{eq:structurelimit}.
 Current and future experiments on weak lensing and $21\,\text{cm}$ experiments~\cite{Kadota:2013iya,Amendola:2012ys} and CMB polarization measurements \cite{Weinberg:2013aya} will likely shed more light on the possible mass range of light scalar DM.  Regarding the quartic, \cite{oai:arXiv.org:1310.6061} previously set an upper bound of $|\lambda_\phi^{\mathrm{eff}} | \lesssim 10^{-74}  \left(\frac{m_\phi}{10^{-18} \eV }\right)^4$ by requiring that the scalar DM be non-relativistic at matter-radiation equality, to an arbitrary degree of $\gamma-1 \sim 10^{-3}$. Scalar dark matter with mass $\sim 10^{-21}$--$10^{-22} \eV$ as a $\mathcal{O}(0.1)$ fraction of the total DM density has previously been proclaimed to resolve some tension between structure formation simulations based on $\Lambda\mathrm{CDM}$ model and observations \cite{Hu:2000ke, Marsh:2013ywa}.\footnote{For a short review on the small-scale astrophysical controversies, namely the cusp-core problem, the missing-satellite problem, and the too-big-to-fail problem, see~\cite{Weinberg:2013aya}.}
We note that this mass range is straddling the first bound of Eq.~\eqref{eq:structurelimit}, and thus provides extra motivation to look for scalar oscillations in the mass range for which the clock comparison tests of Sec.~\ref{sec:sensitivity} are most sensitive.

In Fig.~\ref{fig:cosmobound}, we show the constraints in Eq.~\eqref{eq:structurelimit} on the $m_\phi$--$|\lambda_\phi^{\mathrm{eff}}|$ parameter space in dark and light gray. We also plot the (negative) quartic contribution of Eq.~\eqref{eq:selfrad2} from a $d_{m_t}\sim10^{-6}$ coupling to the top quark mass term in blue, as well as the effective (negative) quartic generated from gravitational effects in black.  For reference, we include the QCD axion mass-quartic relation $\lambda_a \sim - (m_a/\Lambda_3)^4$ in green.  From Fig.~\ref{fig:cosmobound}, we conclude that couplings $d_i$ that can be probed by the experiment we propose in Secs.~\ref{sec:setup}~\&~\ref{sec:sensitivity} can easily be consistent with cosmological and astrophysical constraints. 

In fact, they are complimentary: the structure formation constraint on the mass motivates searches for DM oscillations with sub-year periods, while the constraint on the quartic coupling motivates searches for $d_i$ couplings much smaller than those accessible with current EP experiments.

\subsection{Isocurvature perturbations}
Non-observation of isocurvature fluctuations in the CMB could also constrain very light dilatons, especially in light of recent evidence for $B$-modes from the BICEP2 collaboration \cite{Ade:2014gua}. 
If these $B$-modes are dominantly sourced by primordial gravitational waves, they would suggest a tensor-scalar ratio of $r \approx 0.2$, favoring a high scale of inflation: $H_\text{infl} \approx 1\cdot 10^{14}\GeV$. Their primordial origin is currently under debate due to potential dust contributions suggested in \cite{Flauger:2014qra} and recently measured by the Planck collaboration \cite{Ade:2014xdh}. In the following, we will discuss the potential implications of a high inflationary scale.

Scalar fields which are effectively massless during inflation experience random fluctuations in the field value $A_\text{iso}$ that is isocurvature in nature. The isocurvature flutuations are uncorrelated with the adiabatic density perturbations $A_\text{s}$, while CMB observations require $A_\text{iso} / A_\text{s} \lesssim 0.04$ \cite{Ade:2013uln}. The random isocurvature perturbations are sufficiently suppressed if the field value $\phi_{0,i}$ during inflation is near the Planck scale \cite{Marsh:2014qoa}:
\begin{align} \label{eq:isocurvature}
\kappa \phi_{0,i} \gtrsim 1 \left(\frac{\rho_\phi}{\rho_\text{DM,U}} \right) \left( \frac{r}{0.16} \right)^{1/2} \left(\frac{0.04}{A_\text{iso} / A_\text{s}} \right)^{1/2}
\end{align}
If we neglect anharmonic terms in the potential, the initial amplitude of the field is typically a few orders of magnitude below the Planck scale:
\begin{align} \label{eq:amplitudeinfl}
\kappa \phi_{0,i} \approx 3 \cdot 10^{-3} \left( \frac{10^{-18}\eV}{m_\phi} \right)^{1/4}\left( \frac{\rho_\phi}{\rho_\text{DM,U}} \right)^{1/2}.
\end{align}
Naively, Eqs.~\eqref{eq:isocurvature}~\&~\eqref{eq:amplitudeinfl} together imply that $\phi$ cannot be all of the DM in the universe: 
\begin{align} \label{eq:isocurvatureconstr}
\frac{\rho_\phi}{\rho_\text{DM,U}} \lesssim 10^{-5} \left( \frac{10^{-18}\eV}{m_\phi} \right)^{1/2}.
\end{align}
 However, this constraint is easily evaded by having a large initial field amplitude satisfying Eq.~\eqref{eq:isocurvature}, and a mechanism which dilutes the energy density in $\phi$ after inflation so as to not overclose the universe.  
  One way to transition from a near-Planckian initial amplitude to one consistent with the current DM energy density is through the cosmological attractor mechanism of \cite{Damour:1994zq} discussed in Appendix~\ref{sec:dpmech}. Isocurvature fluctuations could also be small if $\phi$ has a high mass during inflation. For example, the renormalizable coupling to curvature $\xi \phi^2 R$ for $\xi \sim \mathcal{O}(1)$ gives a classical contribution of $\langle R \rangle \simeq 12H_\text{infl}^2 \sim (10^{14.5}\GeV)^2$  to the mass-squared of $\phi$ during inflation, and has only a tiny effect today \cite{Folkerts:2013tua}. Similar in spirit, one could make the mass dependent on the inflaton field value.  We thus conclude that there exist natural solutions to the isocurvature constraint of Eq.~\eqref{eq:isocurvatureconstr} in our dilaton model.  Even if ${\rho_\phi}/{\rho_\text{DM,U}} \sim 1$ were untenable, the sensitivity of our experiment only scales at the square root of the $\phi$ abundance, with large discovery potential even for subdominant components of the DM energy density.

\subsection{Superradiance}
 Precision observations of rotating black holes can constrain ultralight and weakly interacting scalar fields.  In \cite{Arvanitaki:2010sy}, it is argued that light scalars can form gravitational bound states with a black hole if the Compton wavelength of the scalar is comparable to the Schwarzschild radius of the black hole. A rapidly spinning black hole can then lose energy and angular momentum through a variant of the Penrose process---superradiance---causing the occupation numbers of the scalar field in certain bound energy levels to grow exponentially.  When the resulting ``scalar cloud" collapses after it reaches an instability due to gravitational or self-interactions, a large fraction of the angular momentum of the bound black hole system will be lost.  Observations of old, near-extremal black holes can thus exclude the existence of weakly interacting scalars for a range of Compton wavelengths \cite{Arvanitaki:2010sy}.

The spin and mass of near-extremal black holes can be measured to high precision with X-ray spectroscopy~\cite{Reynolds:2013qqa,Reynolds:2013rva,Walton:2012aw}.  In particular, observations of rapidly rotating black holes in the range $10^6 \mathrm{M_{\odot}} \lesssim M_{\mathrm{BH}} \lesssim 10^8 \mathrm{M_{\odot}}$ excludes non-interacting scalars for the mass range $10^{-18.2} \eV \lesssim m_{\phi} \lesssim 10^{-17.6} \eV$ and $10^{-16.7} \eV \lesssim m_{\phi} \lesssim 10^{-16.1} \eV$, while rapidly spinning stellar-mass black hole observations exclude QCD axions in the range $10^{-12.3} \eV \lesssim m_{\phi} \lesssim 10^{-10.8} \eV$ \cite{MashaXinlu:2014}.  However, in the model we are considering, the mass $m_\phi^2$ and quartic $\lambda_\phi^{\mathrm{eff}}$ are two independent parameters, unlike for an axion where the two are related via the $\cos(a/f_a)$ potential.  If the effective quartic is sufficiently large, instabilities in the scalar cloud form early enough to render supperradiance ineffective and evade the afore-mentioned bound.  In Fig.~\ref{fig:cosmobound}, we plot the superradiance exclusion ranges in the $m_\phi$--$|\lambda_\phi^{\mathrm{eff}}|$ plane in yellow.


\section{Discussion}
\label{sec:conclusions}

The unprecedented precision of atomic clocks allows us to probe new parameter space of dark matter with higher-dimensional scalar couplings way beyond the grasp of fifth-force and EP violation searches. The sensitivity varies depending on whether the DM can couple with dimension-five operators to the gluonic field strength through the coupling $d_g$. In particular, if the DM only has couplings to the electromagnetic field strength or behaves like a Yukawa modulus, fifth-force or EP violation tests which are dominated by the gluonic coupling lose their sensitivity and the improvement in the reach of the couplings $d_e$ and $d_{m_i}$ can be in excess of 8 orders of magnitude. 

The detection reach of our proposed experiments depends on the abundance of $\phi$, while fifth-force searches and EP tests are independent of any cosmic abundance constraints. Nevertheless, the misalignment mechanism of DM production suggests that such an assumption is generic. In fact, for most of the parameter space, the initial displacement of the field has to be tuned to some degree in order to avoid overclosure of our universe with DM. Such a tuning is well justified when anthropic considerations are taken into account. Scanning of the initial displacement of the bosonic field can be easily achieved in a long era of inflation, as every Hubble time $H^{-1}$, any scalar field will fluctuate by an amount of order $H$. Due to this random walk, the bosonic field will scan the entirity of its natural range after $N\lesssim \MPl^2 / H^2$ $e$-foldings.

A major concern in our model of an ultralight scalar field coupling to the SM is the issue of naturalness. Naively putting a hard cutoff $\Lambda \sim 10 \TeV$, we find a mass correction to the $\phi$ mass of order:
\begin{align}
\delta m_\phi^2 \sim \frac{(\kappa d_{m_i} m_i \Lambda)^2}{16\pi^2} \sim (10^{-11} \eV )^2 \left(\frac{d_{m_t}}{10^{-6}} \right)^2,
\end{align}
where we have plugged in the top mass $m_t$ to get a numerical estimate. This suggested mass range is several orders of magnitude away from the parameter space probed by our proposed setups. However, this estimate may be too naive. The cutoff could be much lower than $10\TeV$; for the axion, the cutoff is the QCD scale $\Lambda \sim \Lambda_3$, and for a Higgs portal it is the Higgs mass $\Lambda \sim m_h$. Furthermore, in the framework of the string theory landscape, the idea of naturalness is challenged. In this case, the scalar field could be tuned to be light via environmental selection because otherwise it would overclose the universe, or perhaps the smallness of the dilaton mass is correlated with the smallness of the cosmological constant in our vacuum. 

The continued absence of new physics at the TeV scale weakens the motivation for naturalness of the weak scale and the WIMP miracle. In combination with the omnipresent cosmological constant problem, this expands the possibilities for dark matter. At the same time, the rapid progress in fields like clock technology gives rise to new experimental probes of physics beyond the Standard Model, and may suggest a new direction of smaller scale experimental searches away from the high-energy frontier probed by colliders.

\acknowledgments
We would like to thank Masha Baryakhtar, Savas Dimopoulos, Giorgio Gratta, Leo Hollberg, Kiel Howe, Xinlu Huang, Mark Kasevich, John March-Russell, Jeremy Mardon, Maxim Pospelov, Alex Sugarbaker, Yijun Tang, Gabriele Veneziano, Xu Yi, and Tim Wiser for helpful discussions. 
This work is partially supported by ERC grant BSMOXFORD N$^\circ$228169.
Research at Perimeter Institute is supported by the Government of Canada through Industry Canada and by the Province of Ontario through the Ministry of Economic Development \& Innovation.

\appendix

\section{Ultraviolet embedding}\label{sec:dpmech}

Perturbative formulations of string theory predict the existence of a scalar partner to the graviton---the dilaton---which typically has gravitational-strength couplings to ordinary matter. As we have explained in Sec.~\ref{sec:EPtests}, such couplings are excluded quite robustly by non-observations of equivalence principle violations.  Two classes of solutions to this apparent problem exist in the literature: either the dilaton gets a sufficiently high mass (so the dilatonic force is short-range and thus less constrained), or there is some mechanism at work to suppress low-energy couplings to the Standard Model.  A construction of the latter scenario was first attempted in \cite{Damour:1994zq,Damour:1994ya} for a \textit{massless} dilaton, where a cosmological attractor mechanism (dubbed the ``least coupling principle" and first noticed in \cite{Damour:1993id,Damour:1992kf} in the context of scalar-tensor theories of gravity) was responsible for diluting the couplings.

We summarize the final conclusions of \cite{Damour:1994zq}, and discuss the obstacles of generalizations to our model where the dilaton $\phi$ is to be the DM and thus necessarily has a mass (albeit a very small one).  The dynamics of the dilaton interactions are encoded in the lagrangian:
\begin{align} \label{eq:DamPol}
\mathcal{L} \supset V(\phi) + B_i(\kappa \phi)\mathcal{O}_i^\text{SM},
\end{align}
where $\kappa \phi \equiv \kappa \phi \equiv \frac{\sqrt{4\pi}}{\MPl} \phi$ is the field in (modified) Planck units as before and $\mathcal{O}_i^\text{SM}$ is a schematic for SM operators such as $m_i \bar{\psi}_i \psi_i$. We redefine $\phi$ such that $\phi=0$ is the minimum of $V(\phi)$.  
For simplicity, let us first take the assumption of universal $B_i(\kappa \phi) = B(\kappa \phi)$ with a local minimum (another possibility is a runaway direction \cite{Damour:2002mi,Damour:2002nv}). The attractor mechanism is operative in the early stages of the radiation-dominated era, when $H^2 \gg V''(\phi)$ and we can effectively treat the dilaton as massless even in the presence of a non-trivial potential $V(\phi)$. 
As the universe cools down and passes through several mass thresholds of SM particles, $\phi$ is gradually attracted to the minimum $\phi_0^B$ of $B(\kappa \phi)$, to settle at a value $\phi_0^* = \phi_0^B + \Delta \phi_0^*$.  If the curvature $C_i \equiv B_i''(\kappa \phi)$ is sufficiently large, the attraction is efficient and $|\Delta \phi_0^*| \ll \phi_0^B$.
If the dilaton is exactly massless and has no self-interactions ($V(\phi)$ identically zero) as assumed in \cite{Damour:1994zq}, the EP-violating forces are explained to be small because $\phi_0^*$ is so close to $\phi_0^B$ at the present day, with $|d_i| \sim C_i \frac{\Delta \phi_0^*}{\phi_0^*} \ll 1$ in Eq.~\eqref{eq:couplings}.

While the above mechanism is an elegant way to dilute the dilaton couplings, there are several problems with this construction.  The effective potential $V(\phi)$ and dilaton coupling functions $B_i(\kappa \phi)$ are renormalized through loops of dilaton, graviton and matter fields, and perhaps also non-perturbative string dynamics.  
These (incalculable) effects will in general break the assumed universality of the $B_i(\kappa \phi)$ functions; in particular, the minima of the $B_i(\kappa \phi)$ will not coincide, precluding an explanation of the smallness of \textit{most} $d_i$ couplings.  An additional complication arises when the low-energy effective potential $V(\phi)$ is taken into account, even in the case of universal $B_i = B$ with minimum $\phi_0^B$.  Internal dilaton loops generally split the degeneracy of the minima of $V(\phi)$ and $B(\kappa \phi)$, or $\phi_0^B \neq 0$. When the universe cools down to $H^2 \sim V''(\phi)$, the potential will kick in and displace $\phi_0$ away from $\phi_0^B$, undoing the attraction to the $B(\kappa \phi)$ minimum.

We consider the more generic possibility of non-overlapping coupling functions $B_i(\kappa \phi)$ and non-trivial scalar potential $V(\phi)$. In our model, where we assume the dilaton to be a significant fraction of the dark matter produced through vacuum misalignment, we \textit{actually require the field value to be non-zero} at the start of the oscillating phase ($\phi_0^* \neq 0$), because otherwise $\rho_\phi$ would be negligible today.  Assuming the mass term dominates the potential at all times, we can calculate $\phi_0^*$ to be the amplitude of the field when it starts oscillating at $H \sim m_\phi$:
\begin{align} \label{eq:initialamp}
\kappa \phi_0^* \equiv \kappa \phi_0^* \approx 3 \cdot 10^{-3} \left( \frac{10^{-18}\eV}{m_\phi} \right)^{1/4}\left( \frac{\rho_\phi}{\rho_\text{DM}} \right)^{1/2}.
\end{align}
After $\rho_\phi$ redshifts to its present-day value, we would expect $|d_i| \sim C_i \kappa \phi_0^*$ at present, which is typically ruled out for $C_i \sim \mathcal{O}(1)$ and the masses  considered in this paper ($m_\phi < 10^{-15} \eV$), as we showed in Sec.~\ref{sec:EPtests}.  

In summary, the mechanism of \cite{Damour:1994zq} is efficient when the curvature of the $B_i(\kappa \phi)$ functions is large, and can suppress the $d_i$ couplings when the minima of the $B_i(\kappa \phi)$ and $V(\phi)$  are approximately degenerate, which requires functional tuning in Eq.~\eqref{eq:DamPol} unless some symmetry principle is at work. It has been suggested that $S$-duality, with transformations $g_s \rightarrow 1/g_s$ for the string coupling and $\phi \rightarrow - \phi$ as symmetries of the UV theory, could possibly help yield a minimum for all $B_i(\kappa \phi)$ and $V(\phi)$ at $\phi \simeq 0$ without tuning.  For large-curvature $B_i(\kappa \phi)$ functions, the dilaton cannot both be a significant component of DM \textit{and} have small $d_i$ couplings, as Eq.~\eqref{eq:initialamp} implies a non-degeneracy of the $V(\phi)$ and $B_i(\kappa \phi)$ minima in this case.  We thus conclude that dilatonic dark matter models should have \textit{a priori} low-curvature $B_i(\kappa \phi)$ coupling functions in the field space region close to the minimum of the effective potential $V(\phi)$ in order to have small EP-violating effects.  We hope that our above discussion and the potential discovery reach of our proposed experiment in Secs.~\ref{sec:setup}~\&~\ref{sec:sensitivity} spurs future UV-model-building efforts in this direction.

\bibliography{lsdm.bib}

\end{document}